\documentclass[aps,reprint,pre,showpacs]{revtex4-1}

\usepackage{graphics,epsfig}
\usepackage{amsmath,amssymb}
\usepackage{bm}

\begin{document}
\title{Scale coupling and interface pinning effects in the phase-field-crystal model}
\author{Zhi-Feng Huang}
\affiliation{Department of Physics and Astronomy, Wayne State University, 
Detroit, Michigan 48201}
\date{\today}

\begin{abstract}
Effects of scale coupling between mesoscopic slowly-varying envelopes of
liquid-solid profile and the underlying microscopic crystalline structure are 
studied in the phase-field-crystal (PFC) model. Such scale coupling leads to 
nonadiabatic corrections to the PFC amplitude equations, the effect of which
increases strongly with decreasing system temperature below the melting point. 
This nonadiabatic amplitude representation is further coarse-grained for the
derivation of effective sharp-interface equations of motion in the limit of 
small but finite interface thickness. We identify a generalized form of the 
Gibbs-Thomson relation with the incorporation of coupling and pinning effects 
of the crystalline lattice structure. This generalized interface equation can 
be reduced to the form of a driven sine-Gordon equation with KPZ nonlinearity, 
and be combined with other two dynamic equations in the sharp interface limit 
obeying the conservation condition of atomic number density in a liquid-solid 
system. A sample application to the study of crystal layer growth is
given, and the corresponding analytic solutions showing lattice pinning and 
depinning effects and two distinct modes of continuous vs. nucleated growth 
are presented. We also identify the universal scaling behaviors governing the
properties of pinning strength, surface tension, interface kinetic coefficient,
and activation energy of atomic layer growth, which accommodate all range of 
liquid-solid interface thickness and different material elastic modulus.
\end{abstract}


\maketitle

\section{Introduction}
\label{sec:intro}

Continuum theories have been playing a continuously important role
in modeling and understanding a wide range of complex nonequilibrium 
phenomena during materials growth and processing. For the typical example 
of liquid-solid front motion and interface growth, sharp-interface or 
Stefan-type models have been used in early studies to examine various 
solidification phenomena such as dendritic growth and directional 
solidification of either pure systems or eutectic alloys \cite{re:langer80}.
Recent focus has been put on the continuum phase-field approach, which
has become a widely-adopted method in materials modeling not only due
to its computational advantage as compared to atomistic techniques and
also to the complex moving-boundary problems encountered in sharp-interface
models, but also due to its vast applicability for a wide variety of
material phenomena including solidification, phase transformation, alloy 
decomposition, nucleation, defects evolution, nanostructure formation,
etc. \cite{re:elder94,re:karma98,re:elder01,re:muller99,re:kassner01,%
re:granasy04,re:wang10}.

These continuum methods are well formulated for the description of long 
wavelength behavior of a system. To incorporate properties related
to smaller-scale crystalline details which can have significant impact, 
additional assumptions or modifications are required. Examples
include the consideration of lattice anisotropy for surface tension
and kinetics \cite{re:karma98}, and the incorporation of subsidiary
fields describing system elasticity \cite{re:muller99,re:kassner01}, 
plasticity \cite{re:wang10}, or local crystal orientation \cite{re:granasy04} 
in phase-field models. However, effects associated with the discreteness of 
a crystalline system, such as the atomistic feature of lattice growth,
are usually absent due to the nature of continuum description. Efforts
to partially remedy this in some previous studies include, e.g., the adding
of a periodic potential mimicking effects of crystalline lattice along the
growth direction in the continuum modeling of surface roughening transition
\cite{re:chui78,re:nozieres87,re:hwa91,*re:balibar92,*re:mikheev93,*re:rost94,*re:hwa94}, 
although both the form of lattice potential (usually assumed as a sinusoidal 
function as part of the sine-Gordon Hamiltonian) and the associated parameters 
were introduced phenomenologically.

More systematic approaches based on some fundamental microscopic-level theories
are needed for the construction of continuum field models that incorporate 
crystalline/atomistic features. One of the recent advances on this front is the
development of phase-field-crystal (PFC) methods \cite{re:elder02,*re:elder04,re:elder07}, 
in which the structure and dynamics of a solid system are described by a continuum 
local atomic density field that is spatially periodic and of atomistic resolution;
thus the small length scale of crystalline lattice structure is intrinsically
built into the continuum field description with diffusive dynamic time scales.
Both free energy functionals and dynamics of the PFC models can be derived from 
atomic-scale theory through classical density functional theory of freezing (CDFT)
and the corresponding dynamic theory (DDFT), for both single-component and alloy 
systems \cite{re:elder07,re:huang10b,re:greenwood10,*re:greenwood11b,re:teeffelen09,%
re:jaatinen09}. Properties associated with crystalline nature of the system, such
as elasticity, plasticity, multiple grain orientations, crystal symmetries and
anisotropy, are then naturally included, with no additional phenomenological 
assumptions needed as compared to conventional continuum field theories. This 
advantage has been verified in a large variety of applications of PFC, ranging from
structural, compositional, to nanoscale phenomena for both solid materials 
\cite{re:elder02,*re:elder04,re:elder07,re:huang10b,re:greenwood10,*re:greenwood11b,%
re:jaatinen09,re:huang08,*re:huang10,re:wu09,re:spatschek10,re:muralidharan10,%
re:berry11,re:elder12} and soft matters \cite{re:teeffelen09,re:wittkowski11}.

An important feature of the PFC methodology is the multiple scale description it
provides, as can be seen from its amplitude representation. The system dynamics 
is described by the behavior of ``slow''-scale (mesoscopic) amplitudes/envelopes 
of the underlying crystalline lattice, as a result of the amplitude expansion of
PFC density field in either pure liquid-solid systems
\cite{re:goldenfeld05,*re:athreya06,re:yeon10} or binary alloys 
\cite{re:elder10a,re:huang10b,re:spatschek10}. Note that in these amplitude 
equation studies, although most lattice effects have been incorporated in the 
variation of complex amplitudes (mainly via their phase dynamics 
\cite{re:huang08,*re:huang10}), the spatial scales of the \textit{mesoscopic} 
amplitudes and the \textit{microscopic} lattice structure are assumed to be 
separated (i.e., the assumption of ``adiabatic'' expansion). 
However, this assumption only holds in the region of slowly varying
density profile either close to the bulk state or for diffuse interfaces, and
hence is valid only at high enough system temperature. In low or intermediate
temperature regime showing sharp liquid-solid or grain-grain interfaces, 
amplitude variation around the interface would be of order close to the 
lattice periodicity; thus the two scales of amplitudes vs. lattice can no longer
be separated, resulting in the ``nonadiabatic'' effect due to their coupling 
and interaction. Such scale coupling leads to an important effect of lattice 
pinning that plays a pivotal role in material growth and evolution, as first
discussed by Pomeau \cite{re:pomeau86} and later demonstrated in the phenomena
of fluid convection and pattern formation \cite{re:bensimon88b,re:boyer02,*re:boyer02b}.
To our knowledge, these scale coupling effects have not been addressed explicitly
in all previous phase-field and PFC studies of solidification and crystal growth.

In this paper we aim to identify these coupling effects between \textit{mesoscale} 
structural amplitudes and the underlying \textit{microscopic} spatial scale of 
crystalline structure, via deriving the nonadiabatic amplitude representation of 
the PFC model. What we study here is the simplest PFC system: two-dimensional 
(2D), single-component, and of hexagonal crystalline symmetry, as our main focus
is on examining the fundamental aspects of scale coupling and bridging that are
missing in previous research, and also on further completing the multi-scale features
of the PFC methodology. The explicit expression of the resulting pinning
force during liquid-solid interface motion, and also its scaling behavior with 
respect to the interface thickness, are determined in this work, through the
application of sharp/thin interface approach (given finite interface thickness) to 
the amplitude equations. This leads to a new set of interface equations of motion,
in particular a generalized Gibbs-Thomson relation that incorporates the pinning 
term and also its reduced form of a driven sine-Gordon equation. The pinning of 
the interface to the underlying crystalline lattice structure, and the associated
nonactivated vs. nucleated growth modes, can be determined from analytic 
solutions of the interface equations for the case of planar layer growth.

\section{Nonadiabatic coupling in amplitude equations}
\label{sec:ampl}

In the PFC model for single-component systems, the dynamics of a rescaled atomic 
number density field $n({\bm r},t)$ is described in a dimensionless form 
\cite{re:elder02,*re:elder04,re:elder07,re:huang10b}
\begin{equation}
\partial n / \partial t = \nabla^2 \left [ -\epsilon n + (\nabla^2 +
  q_0^2)^2 n - g n^2 +n^3 \right ] + {\bm \nabla} \cdot {\bm \eta},
\label{eq:pfc}
\end{equation}
where $\epsilon$ measures the temperature distance from the melting point,
$g=(3/B^x)^{1/2}/2$ with $B^x$ proportional to the 
bulk modulus, and we have $q_0=1$ after rescaling over a length scale $R$ of 
lattice spacing. The noise field ${\bm \eta}$ has zero mean and obeys the 
correlations
\begin{equation}
\langle \eta^{\alpha}({\bm r},t) \eta^{\beta}({\bm r'},t') \rangle =
2\Gamma_0 k_BT \delta ({\bm r} - {\bm r'}) \delta (t-t') \delta^{\alpha\beta}
\label{eq:noise}
\end{equation}
with $\alpha, \beta = x, y, z$, where $\Gamma_0$ is a rescaled constant
depending on $B^x$ and $R$ \cite{re:huang10b}, and $T$ is the system 
temperature.

To derive the corresponding 2D amplitude equations in the limit of small
$\epsilon$, we need to first distinguish the ``slow'' spatial and temporal 
scales for the amplitudes/envelopes of the structural profile, i.e.,
\begin{equation}
X = \epsilon^{1/2} x, \qquad Y = \epsilon^{1/2} y, \qquad T = \epsilon t,
\label{eq:scaling}
\end{equation}
from the ``fast'' scales $(x,y,t)$ of the underlying hexagonal
crystalline structure. We then expand the PFC model equation (\ref{eq:pfc})
based on this scale separation and also on a hybrid approach combining the
standard multiple-scale expansion\cite{re:manneville90,re:cross93} and 
the idea of ``Quick-and-Dirty'' renormalization group method 
\cite{re:goldenfeld05,*re:athreya06} (see Ref. \cite{re:huang10b} for
details). To incorporate the coupling between these ``slow'' and ``fast''
scales, which leads to nonadiabatic corrections to the amplitude
equations, we use an approach based on that given in Refs. 
\cite{re:bensimon88b,re:boyer02,*re:boyer02b} which address front motion 
and locking in periodic pattern formation during fluid convection.

Following the steps of standard multiple-scale analysis 
\cite{re:manneville90,re:cross93}, the atomic density field $n$
can be expanded as
\begin{equation}
n = n_0(X,Y,T) + \sum\limits_{j=1}^{3} A_j(X,Y,T)
e^{i {\bm q}_j^0 \cdot {\bm r}} + \text{c.c.},
\label{eq:n_expan}
\end{equation}
where ${\bm q}_j^0$ are the three basic wave vectors for 2D hexagonal
structure (i.e., the 3 ``fast''-scale base modes)
\begin{equation}
{\bm q_1^0} = -q_0 \left ( \frac{\sqrt{3}}{2} \hat{x} + \frac{\hat{y}}{2} 
\right ),  {\bm q_2^0} = q_0 \hat{y}, 
{\bm q_3^0} = q_0 \left ( \frac{\sqrt{3}}{2} \hat{x} - \frac{\hat{y}}{2} 
\right ),
\label{eq:qj0}
\end{equation}
and the slow scaled fields, including $A_j$ (complex amplitudes
of mode ${\bm q}_j^0$) and $n_0$ (real amplitude of the zero wavenumber 
neutral mode as a result of PFC conserved dynamics), are represented as 
power series of $\epsilon$:
$A_j = \epsilon^{1/2} A_j^{(1/2)} + \epsilon A_j^{(1)} + \epsilon^{3/2}
A_j^{(3/2)} + \cdots = \sum_{m=1}^{\infty} \epsilon^{m/2} A_j^{(m/2)}$,
$n_0 = \epsilon^{1/2} n_0^{(1/2)} + \epsilon n_0^{(1)} + \epsilon^{3/2}
n_0^{(3/2)} + \cdots = \sum_{m=1}^{\infty} \epsilon^{m/2} n_0^{(m/2)}$.
Note that in Eq. (\ref{eq:n_expan}) higher harmonic terms have been
neglected.

From Eqs. (\ref{eq:scaling}) and (\ref{eq:n_expan}) as well as the
substitutions $\partial_{x(y)} \rightarrow \partial_{x(y)} + 
\epsilon^{1/2} \partial_{X(Y)}$ and
$\partial_t \rightarrow \epsilon \partial_T$, we obtain the
following expansion for the PFC equation (\ref{eq:pfc}) in the
absence of noise:
\begin{widetext}
\begin{eqnarray}
\left [ \mathcal{L} n + \nabla^2 \left (g n^2 -n^3 \right ) \right ]_s 
&& = \epsilon \partial_T n_0 - \epsilon \nabla_s^2 
\frac{\delta \mathcal{F}_s}{\delta n_0} + \sum_{j=1}^3 \left [ \left ( 
\epsilon \partial_T A_j - \mathcal{L}_j^s \frac{\delta \mathcal{F}_s}
{\delta A_j^*} \right ) e^{i {\bm q}_j^0 \cdot {\bm r}} + \text{c.c.}
\right ] \nonumber\\
&& + \left [ f_{p_{11}} e^{2i {\bm q}_1^0 \cdot {\bm r}} 
+ f_{p_2} e^{2i {\bm q}_2^0 \cdot {\bm r}}
+ f_{p_{33}} e^{2i {\bm q}_3^0 \cdot {\bm r}}
+ f_{p_1} e^{i ({\bm q}_1^0 - {\bm q}_2^0) \cdot {\bm r}}
+ f_{p_0} e^{i ({\bm q}_1^0 - {\bm q}_3^0) \cdot {\bm r}}
+ f_{p_3} e^{i ({\bm q}_3^0 - {\bm q}_2^0) \cdot {\bm r}}
+ \text{c.c.} \right ] \nonumber\\
&& + \sum_{j\neq k =1}^3 f_{p_{jk}} e^{i (2{\bm q}_j^0 - {\bm q}_k^0) 
\cdot {\bm r}} + 9 q_0^2 \sum_{j=1}^3 A_j^3 e^{3i {\bm q}_j^0 \cdot {\bm r}}
+ \text{c.c.},
\label{eq:pfc_s}
\end{eqnarray}
\end{widetext}
where $\mathcal{L} = \partial_t + \epsilon \nabla^2 - \nabla^2 
(\nabla^2 + q_0^2)^2$ is the linear operator in PFC, $[ \cdots ]_s$ refers 
to the slow-scale expansion to all orders of $\epsilon$, and
$\bm{\nabla}_s = (\partial_X, \partial_Y)$, $\nabla_s^2 = \partial_X^2
+ \partial_Y^2$, and $\mathcal{L}_j^s = \epsilon \nabla_s^2 + \epsilon^{1/2}
\left ( 2i {\bm q}_j^0 \cdot {\bm \nabla}_s \right ) - q_0^2$ are slow 
operators. In Eq. (\ref{eq:pfc_s}), $\mathcal{F}_s$ is the slow-scale
correspondence of the effective free energy $\mathcal{F}$ given below
[with $(\nabla^2 + 2i {\bm q}_j^0 \cdot {\bm \nabla})$ replaced by
$(\mathcal{L}_j^s + q_0^2)$ and $(\nabla^2+q_0^2)$ replaced by
$(\epsilon \nabla_s^2+q_0^2)$]:
\begin{eqnarray}
\mathcal{F}=&& \int d \bm{r} \left \{ (-\epsilon + 3 n_0^2 - 2gn_0)
\sum_{j=1}^3 |A_j|^2 \right. \nonumber\\
&& + \sum_{j=1}^3 \left | \left ( \nabla^2 + 2i {\bm q}_j^0 
\cdot {\bm \nabla} \right ) A_j \right |^2 + \frac{3}{2} \sum_{j=1}^3 |A_j|^4
\nonumber\\
&& + (6n_0-2g) \left ( \prod_{j=1}^3 A_j + \text{c.c.} \right )
+ 6 \sum_{j<k} |A_j|^2 |A_k|^2 \nonumber\\
&& \left. - \frac{1}{2} \epsilon n_0^2 +
  \frac{1}{2} \left [ \left ( \nabla^2 + q_0^2 \right ) n_0 \right ]^2
  - \frac{1}{3} g n_0^3 + \frac{1}{4} n_0^4 \right \},
\label{eq:F}
\end{eqnarray}
which is the same as the previous amplitude expansion result
\cite{re:yeon10,re:huang08,*re:huang10,re:huang10b}; also for other 
variables $f_{p_{jk}}$ ($j,k=1,2,3$) and $f_{p_i}$ ($i=0,...,3$) related 
to higher harmonics,
\begin{eqnarray}
f_{p_1} &&= 3q_0^2 \left [ (6n_0-2g) A_1A_2^* + 3 \left ( A_1^2A_3 
+ {A_2^*}^2A_3^* \right ) \right ], \nonumber\\
f_{p_2} &&= 4q_0^2 \left [ (3n_0-g) A_2^2 + 6A_1^*A_2A_3^* \right ],
\nonumber\\
f_{p_3} &&= 3q_0^2 \left [ (6n_0-2g) A_2^*A_3 + 3 \left ( A_1A_3^2 
+ A_1^*{A_2^*}^2 \right ) \right ], \nonumber\\
f_{p_0} &&= 3q_0^2 \left [ (6n_0-2g) A_1A_3^* + 3 \left ( A_1^2A_2 
+ A_2^*{A_3^*}^2 \right ) \right ], \nonumber\\
f_{p_{11}} &&= 4q_0^2 \left [ (3n_0-g) A_1^2 + 6A_1A_2^*A_3^* \right ],
\nonumber\\
f_{p_{33}} &&= 4q_0^2 \left [ (3n_0-g) A_3^2 + 6A_1^*A_2^*A_3 \right ],
\nonumber\\
f_{p_{jk}} &&= 21q_0^2 A_j^2 A_k^* ~~(j \neq k).
\label{eq:fp}
\end{eqnarray}

\begin{figure}
\centerline{
\includegraphics[width=0.35\textwidth]{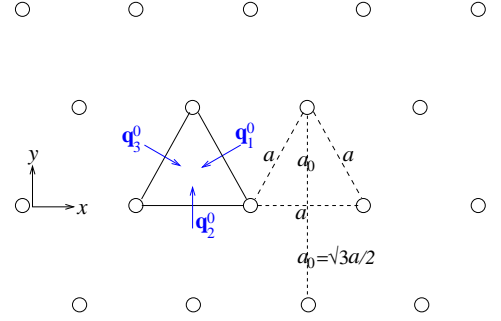}}
\caption{(Color online) Schematic of a hexagonal/triangular lattice, 
with atomic lattice spacing $a=a_0/(\sqrt{3}/2)$, $a_0=2\pi /q_0$, and 
3 basic wave vectors ${\bm q}_j^0$ ($j=1,2,3$) expressed in 
Eq. (\ref{eq:qj0}).}
\label{fig:hex}
\end{figure}

As in the hybrid method developed in Ref. \cite{re:huang10b}, the amplitude
equations governing $A_j$ and $n_0$ can be derived from the integration
of Eq. (\ref{eq:pfc_s}) over eigenmodes $\{ e^{-i {\bm q}_j^0 \cdot {\bm r}}, 
1 \}$, i.e.,
\begin{eqnarray}
\textstyle
& \int_{x}^{x+\lambda_x} \frac{dx'}{\lambda_x} \int_{y}^{y+\lambda_y} 
\frac{dy'}{\lambda_y} \left [ \mathcal{L} n + \nabla^2 
(g n^2 -n^3) \right ]_s e^{-i {\bm q}_j^0 \cdot {\bm r'}} =0, &
\nonumber\\
& \int_{x}^{x+\lambda_x} \frac{dx'}{\lambda_x} \int_{y}^{y+\lambda_y} 
\frac{dy'}{\lambda_y} \left [ \mathcal{L} n + \nabla^2 
(g n^2 -n^3) \right ]_s =0, &
\label{eq:int_ampl}
\end{eqnarray}
where $\lambda_x=a=a_0/(\sqrt{3}/2)$ and $\lambda_y=\sqrt{3}a=2a_0$ 
(with $a$ the atomic lattice spacing for the hexagonal/triangular 
structure and $a_0=2\pi /q_0$, as illustrated in Fig. \ref{fig:hex}), 
which are the atomic spatial periods along the $x$ and $y$ directions 
respectively. Note that Eq. (\ref{eq:int_ampl}) can be also viewed as 
the combination of the solvability conditions obtained at all different 
orders of $\epsilon$ in multiple-scale expansion \cite{re:huang10b}.

In the limit of $\epsilon \rightarrow 0$, i.e., close to the melting
temperature, the spatial variation of amplitudes $A_j(X,Y,T)$ and 
$n_0(X,Y,T)$ is of much larger scale compared to the atomic lattice
variation scales $\lambda_x$ and $\lambda_y$. Thus ``slow'' and 
``fast'' length scales in the integral of Eq. (\ref{eq:int_ampl}) can 
be separated as in standard multiple-scale analysis (i.e., $(X,Y)$ 
and $(x,y)$ be treated as independent variables), leading to the
Ginzburg-Landau-type amplitude equations obtained in previous studies
\cite{re:yeon10,re:huang08}:
\begin{equation}
\partial A_j / \partial t = \mathcal{L}_j \frac{\delta \mathcal{F}}
{\delta A_j^*} \simeq -q_0^2 \frac{\delta \mathcal{F}}{\delta A_j^*}, \quad 
\partial n_0 / \partial t = \nabla^2 \frac{\delta \mathcal{F}}{\delta n_0}.
\label{eq:ampl}
\end{equation}
Note that to derive Eq. (\ref{eq:ampl}) the long-wavelength approximation 
$\mathcal{L}_j = \nabla^2 + 2i {\bm q}_j^0 \cdot {\bm \nabla} - q_0^2 
\simeq -q_0^2$ has been used as before.

However, such assumption of scale separation would not hold when 
$\epsilon$ is of larger value (still small but finite, corresponding 
to low/moderate material temperature far enough from the melting point).
Although the amplitudes/envelopes still vary slowly in the bulk,
the interface, either between liquid and solid states or between different 
grains, could be thin or sharp, with its width comparable to ``fast''
lattice scales (e.g., of few lattice spacings). Thus functions of
$A_j$ and $n_0$ in Eq. (\ref{eq:pfc_s}) can no longer be decoupled from 
the atomic-scale oscillatory terms $e^{i (l {\bm q}_j^0 - m {\bm q}_k^0)
\cdot {\bm r}}$ (with $l$ and $m$ integers and $j,k=1,2,3$) in the
integration of $\int_{x}^{x+\lambda_x} dx' \int_{y}^{y+\lambda_y} dy'$ in 
Eq. (\ref{eq:int_ampl}). Using an approximation similar to that in
Ref. \cite{re:boyer02,*re:boyer02b}, we only keep the lowest-order
coupling terms, i.e., terms coupled to lowest modes (with largest 
atomic lengths) along both $x$ and $y$ directions, including 
$e^{i {\bm q}_j^0 \cdot {\bm r}}$, corresponding to atomic layer spacing
$\sqrt{3} a/2 = a_0$, and $e^{i ({\bm q}_j^0 - {\bm q}_k^0)_{j \neq k}
\cdot {\bm r}}$, with atomic layer spacing $a/2 = a_0/\sqrt{3}$.
Couplings to higher modes, such as $e^{2i {\bm q}_j^0 \cdot {\bm r}}$ (with 
length scale $a_0/2$), $e^{i (2{\bm q}_j^0 - {\bm q}_k^0)_{j \neq k} \cdot {\bm r}}$
(with length scale $a_0/\sqrt{7}$), etc., are neglected. The following
scale-coupled amplitude equations can then be derived from 
Eqs. (\ref{eq:int_ampl}) and (\ref{eq:pfc_s}):

\begin{widetext}
\begin{eqnarray}
\partial A_1 / \partial t =&& -q_0^2 \frac{\delta \mathcal{F}}
{\delta A_1^*} \nonumber\\
&& -\frac{1}{\lambda_x \lambda_y} \int_{x}^{x+\lambda_x} dx' 
\int_{y}^{y+\lambda_y} dy' \left [ f_{p_1} e^{-i q_0 y'}
+ f_{p_{11}} e^{-i q_0 \left ( \frac{\sqrt{3}}{2} x' + \frac{1}{2} y' \right )}
+ f_{p_0} e^{i q_0 \left ( -\frac{\sqrt{3}}{2} x' + \frac{1}{2} y' \right )}
\right. \nonumber\\
&& \left. + f_{p_{13}} e^{-i q_0 \sqrt{3} x'}
+ f_{p_{12}} e^{-i q_0 \left ( \frac{\sqrt{3}}{2} x' + \frac{3}{2} y' \right )}
+ f^*_{p_2} e^{i q_0 \left ( \frac{\sqrt{3}}{2} x' - \frac{3}{2} y' \right )}
+ f^*_{p_{33}} e^{i q_0 \left ( -\frac{\sqrt{3}}{2} x' + \frac{3}{2} y' \right )}
\right ] + \eta_1, \label{eq:ampl1}\\
\partial A_2 / \partial t =&& -q_0^2 \frac{\delta \mathcal{F}}
{\delta A_2^*} \nonumber\\
&& -\frac{1}{\lambda_x \lambda_y} \int_{x}^{x+\lambda_x} dx' 
\int_{y}^{y+\lambda_y} dy' \left [ f_{p_2} e^{i q_0 y'}
+ f^*_{p_1} e^{i q_0 \left ( \frac{\sqrt{3}}{2} x' + \frac{1}{2} y' \right )}
+ f^*_{p_3} e^{i q_0 \left ( -\frac{\sqrt{3}}{2} x' + \frac{1}{2} y' \right )}
\right. \nonumber\\
&& \left. + f^*_{p_{11}} e^{i q_0 \sqrt{3} x'} + f^*_{p_{33}} e^{-i q_0 \sqrt{3} x'}
+ f_{p_{21}} e^{i q_0 \left ( \frac{\sqrt{3}}{2} x' + \frac{3}{2} y' \right )}
+ f_{p_{23}} e^{i q_0 \left ( -\frac{\sqrt{3}}{2} x' + \frac{3}{2} y' \right )}
\right ] + \eta_2, \label{eq:ampl2}\\
\partial A_3 / \partial t =&& -q_0^2 \frac{\delta \mathcal{F}}
{\delta A_3^*} \nonumber\\
&& -\frac{1}{\lambda_x \lambda_y} \int_{x}^{x+\lambda_x} dx' 
\int_{y}^{y+\lambda_y} dy' \left [ f_{p_3} e^{-i q_0 y'}
+ f^*_{p_0} e^{i q_0 \left ( \frac{\sqrt{3}}{2} x' + \frac{1}{2} y' \right )}
+ f_{p_{33}} e^{i q_0 \left ( \frac{\sqrt{3}}{2} x' - \frac{1}{2} y' \right )}
\right. \nonumber\\
&& \left. + f_{p_{31}} e^{i q_0 \sqrt{3} x'}
+ f^*_{p_{11}} e^{i q_0 \left ( \frac{\sqrt{3}}{2} x' + \frac{3}{2} y' \right )}
+ f^*_{p_2} e^{-i q_0 \left ( \frac{\sqrt{3}}{2} x' + \frac{3}{2} y' \right )}
+ f_{p_{32}} e^{i q_0 \left ( \frac{\sqrt{3}}{2} x' - \frac{3}{2} y' \right )}
\right ] + \eta_3, \label{eq:ampl3}\\
\partial n_0 / \partial t =&& \nabla^2 \frac{\delta \mathcal{F}}{\delta n_0}
\nonumber\\
&& -\frac{1}{\lambda_x \lambda_y} \int_{x}^{x+\lambda_x} dx' 
\int_{y}^{y+\lambda_y} dy' \left [ f^*_{p_0} e^{i q_0 \sqrt{3} x'}
+ f^*_{p_1} e^{i q_0 \left ( \frac{\sqrt{3}}{2} x' + \frac{3}{2} y' \right )}
+ f_{p_3} e^{i q_0 \left ( \frac{\sqrt{3}}{2} x' - \frac{3}{2} y' \right )}
+ \text{c.c.} \right ] + \bm{\nabla} \cdot \bm{\eta}_0,
\label{eq:ampl0}
\end{eqnarray}
\end{widetext}
where a projection procedure has been applied to address the noise
term of the PFC Eqs. (\ref{eq:pfc}) and (\ref{eq:noise}) \cite{re:huang10b}, 
leading to zero mean of noise amplitudes $\eta_j$ and $\bm{\eta}_0$, 
as well as the  correlations $\langle \eta_i \eta_j \rangle = \langle 
{\bm \eta}_0 \eta_j \rangle = \langle {\bm \eta}_0 \eta_j^* \rangle =0$
and
\begin{eqnarray}
& \langle \eta_i \eta_j^* \rangle = 2 \vartheta_i q_0^2 \Gamma_0 k_BT
\delta ({\bm r} - {\bm r'}) \delta (t-t') \delta_{ij}, & \nonumber\\
& \langle \eta_0^{\alpha} \eta_0^{\beta} \rangle = 2 \vartheta_0
\Gamma_0 k_BT \delta ({\bm r} - {\bm r'}) \delta (t-t') \delta^{\alpha\beta},&
\label{eq:eta}
\end{eqnarray}
with $i,j=1,2,3$, $\alpha,\beta=x,y$, and $\vartheta_i = \vartheta_0 =1/7$
if assuming equal contribution from all eigenmodes 
$\{ e^{i {\bm q}_j^0 \cdot {\bm r}}, 1 \}$.
In the above generalized amplitude equations (\ref{eq:ampl1})--(\ref{eq:ampl0}),
the integration terms explicitly yield the coupling between ``slow''
(for structural amplitudes or envelopes) and ``fast'' (for atomic lattice
variations) spatial scales, that is, the \textit{nonadiabatic corrections}.
The first 3 coupling terms in each of Eqs. (\ref{eq:ampl1})--(\ref{eq:ampl3}) 
for the dynamics of complex amplitudes $A_j$ are associated with lattice modes
$e^{i {\bm q}_j^0 \cdot {\bm r}}$ of length scale $a_0$ ($=\sqrt{3} a/2$), while 
the other 4 terms correspond to the coupling to 
$e^{i ({\bm q}_j^0 - {\bm q}_k^0)_{j \neq k} \cdot {\bm r}}$ modes with length scale 
$a_0/\sqrt{3}$ ($= a/2$). The nonadiabatic effect for $n_0$ dynamics is 
weaker, with only couplings to the $a_0/\sqrt{3}$ length scale 
given in Eq. (\ref{eq:ampl0}). Note that in the bulk state of single
crystal or homogeneous liquid, the amplitude functions $f_{p_i}$, 
$f_{p_{jk}}$ $\sim$ constants, and hence all the integrals in 
Eqs. (\ref{eq:ampl1})--(\ref{eq:ampl3}) are equal to zero; we can then
recover the original amplitude equations (\ref{eq:ampl}) without any 
nonadiabatic coupling, as expected.

\section{Interface equations of motion with lattice pinning}
\label{sec:int_eqs}

To illustrate the important effects of scale coupling identified in
above nonadiabatic amplitude equations, here we consider a system
of coexisting liquid and solid phases, with the average interface 
normal direction pointed along $\hat{y}$. The extension of our 
derivation and results to other interface orientations is straightforward.

In this case, the ``slow'' and ``fast'' scales parallel to the interface 
can be well separated, and to lowest-order approximation the amplitude
equations (\ref{eq:ampl1})--(\ref{eq:ampl3}) are rewritten as
\begin{eqnarray}
\partial A_j / \partial t &=& -q_0^2 \frac{\delta \mathcal{F}}
{\delta A_j^*} -\frac{1}{\lambda_y} \int_{y}^{y+\lambda_y} dy' 
f_{p_j} e^{\mp i q_0 y'} + \eta_j \nonumber\\
&=& -q_0^2 \left [ \left ( \nabla^2 + 2i \bm{q}_j^0 \cdot \bm{\nabla}
\right )^2 A_j + \frac{\partial f}{\partial A_j^*} \right ] \nonumber\\
&& -\frac{1}{\lambda_y} \int_{y}^{y+\lambda_y} dy' 
f_{p_j} e^{\mp i q_0 y'} + \eta_j, \label{eq:ampl_Aj}\\
\partial n_0 / \partial t &=& \nabla^2 \frac{\delta \mathcal{F}}
{\delta n_0} + \bm{\nabla} \cdot \bm{\eta}_0
= \nabla^2 \mu + \bm{\nabla} \cdot \bm{\eta}_0 \nonumber\\
&=& \nabla^2 \left [ \left ( \nabla^4 + 2q_0^2 \nabla^2 \right ) n_0 
+ \frac{\partial f}{\partial n_0} \right ]
+ \bm{\nabla} \cdot \bm{\eta}_0, \label{eq:ampl_n0}
\end{eqnarray}
($e^{- i q_0 y'}$: for $A_1$ and $A_3$; $e^{+i q_0 y'}$: for $A_2$), where 
$\mu = \delta \mathcal{F} / \delta n_0$ is a chemical potential of the 
system, and $f$ is the bulk free energy density given from Eq. (\ref{eq:F}):
\begin{eqnarray}
f=&& (-\epsilon + 3 n_0^2 - 2gn_0) \sum_{j=1}^3 |A_j|^2
+ \frac{3}{2} \sum_{j=1}^3 |A_j|^4
\nonumber\\
&& + (6n_0-2g) \left ( \prod_{j=1}^3 A_j + \text{c.c.} \right )
+ 6 \sum_{j<k} |A_j|^2 |A_k|^2 \nonumber\\
&& + \frac{1}{2} \left ( -\epsilon + q_0^4 \right ) n_0^2
  - \frac{1}{3} g n_0^3 + \frac{1}{4} n_0^4.
\label{eq:f}
\end{eqnarray}

To derive the corresponding equations of motion for the interface with
finite thickness $\xi$ (i.e., in the sharp/thin interface limit 
\cite{re:karma98,re:elder01}), we follow the general approach developed
by Elder \textit{et al.} \cite{re:elder01} with the use of projection 
operator method. In this approach, a small parameter 
$\varepsilon$ is introduced, which represents the role of the interface
P\'eclet number, and the system is partitioned into two regions: an
inner region around the interface, defined by $-\zeta < u < +\zeta$,
and an outer region far from the interface (i.e., $|u| > \zeta$), 
where length $\zeta$ scales as $1 \ll \zeta/\xi \ll \varepsilon^{-1}$,
and $u$ is the component of a local curvilinear coordinate in the
interface normal direction. In this curvilinear coordinate $(u,s)$,
the two orthogonal unit vectors are defined as $\hat{n} = \hat{x} 
\sin \theta + \hat{y} \cos \theta$ (local normal of the interface) and
$\hat{t} = \partial \hat{n} / \partial \theta$ (tangent to the interface),
where $\theta$ is the angle between $\hat{n}$ and the $y$ axis; also
\begin{eqnarray}
&& \bm{\nabla} = \hat{n} \partial_u + \hat{t} \frac{\partial_s}{1+u\kappa},
\nonumber\\
&& \nabla^2 = \partial_u^2 + \frac{\kappa}{1+u\kappa} \partial_u
+\frac{\partial_s^2}{(1+u\kappa)^2} - \frac{u\partial_s\kappa}
{(1+u\kappa)^3} \partial_s, \label{eq:gradient_us}
\end{eqnarray}
with the local curvature $\kappa=\bm{\nabla} \cdot \hat{n}
=\partial \theta / \partial s$.

\subsection{Outer equations}

In the \textit{outer region} which is far enough from the interface and 
close to the bulk states, the scale coupling term in Eq. (\ref{eq:ampl_Aj})
can be neglected, and the slowly varying amplitude fields $A_j$ and
$n_0$ depend on rescaled spatial variables $(\varepsilon u/\xi, 
\varepsilon s/\xi)$ and rescaled time $\varepsilon^2 t$. Expanding
the outer solution of amplitudes in powers of $\varepsilon$, i.e.,
\begin{eqnarray}
&A_j^{\text{out}} = {A_j^0}^{\text{out}} + \varepsilon {\tilde{A}_j}^{\text{out}}
+\cdots,
n_0^{\text{out}} = {n_0^0}^{\text{out}} + \varepsilon {\tilde{n}_0}^{\text{out}}
+\cdots,& \nonumber\\
&\mu^{\text{out}} = \mu_0^{\text{out}} + \varepsilon \tilde{\mu}_1^{\text{out}}
+\cdots,& \label{eq:expan_out}
\end{eqnarray}
substituting them into Eqs. (\ref{eq:ampl_Aj}) and (\ref{eq:ampl_n0}),
and using the rescaling given above, we find that at $\mathcal{O}(1)$,
\begin{equation}
\left. \frac{\partial f}{\partial A_j^*} \right |_0^{\text{out}}=0, \quad
\partial {n_0^0}^{\text{out}} / \partial t = \nabla^2 
\left. \frac{\partial f}{\partial n_0} \right |_0^{\text{out}}
= \nabla^2 \mu_0^{\text{out}},
\label{eq:out_0}
\end{equation}
and at $\mathcal{O}(\varepsilon)$,
\begin{equation}
\left. \frac{\partial f}{\partial A_j^*} \right |_1^{\text{out}}=0, \quad
\partial {\tilde{n}_0}^{\text{out}} / \partial t = \nabla^2 
\left. \frac{\partial f}{\partial n_0} \right |_1^{\text{out}}
= \nabla^2 \tilde{\mu}_1^{\text{out}},
\label{eq:out_1}
\end{equation}
where ``$|_0^{\text{out}}$'' refers to replacing $(A_j,n_0)$ by 
$({A_j^0}^{\text{out}},{n_0^0}^{\text{out}})$ in the derivative
$\partial f / \partial A_j^*$ or $\partial f / \partial n_0$,
and ``$|_1^{\text{out}}$'' refers to the corresponding results up to
1st order of ${\tilde{A}_j}^{\text{out}}$ and ${\tilde{n}_0}^{\text{out}}$. 
Note that if assuming the system to be not far from a 
liquid-solid equilibrium state, Eq. (\ref{eq:out_0}) of 
$\mathcal{O}(1)$ yields the bulk equilibrium solutions of the uniform 
liquid ($u=+\infty$) or solid ($u=-\infty$) state
\begin{equation}
{A_j^0}^{\text{out}}(u) \equiv A_j^0(\pm \infty), \quad
{n_0^0}^{\text{out}}(u) \equiv n_0^0(\pm \infty),
\end{equation}
with the corresponding equilibrium chemical potential 
$\mu_{\text{eq}} = \mu_0^{\text{out}} = (\partial f / \partial n_0)|_0^{\text{out}}$.

\subsection{Inner expansion and lattice coupling effect}
\label{sec:inner}

For the \textit{inner region} ($-\zeta < u < +\zeta$), the amplitudes 
and chemical potential can be also expanded as
\begin{eqnarray}
& A_j^{\text{in}} = A_j^0 + \varepsilon \tilde{A}_j +\cdots, \quad 
n_0^{\text{in}} = n_0^0 + \varepsilon \tilde{n}_0 +\cdots, & \nonumber\\
&\mu^{\text{in}} = \mu_0 + \varepsilon \tilde{\mu}_1 +\cdots.&
\label{eq:expan_in}
\end{eqnarray}
Due to the presence of interface at $u=0$, the amplitudes are expected to
vary rapidly along the normal direction $\hat{n}$ but slowly along the
arclength $s$ of the interface, leading to the rescaling $(U,S) = 
(u/\xi, \varepsilon s/\xi)$. Considering small interface fluctuations 
and noise amplitude, we assume that $\kappa = \varepsilon
\tilde{\kappa}/\xi$, $\theta = \varepsilon \tilde{\theta}$, 
$\eta_j = \varepsilon \tilde{\eta}_j$, and
$\bm{\eta}_0 = \varepsilon \tilde{\bm{\eta}}_0$. Thus from 
Eq. (\ref{eq:gradient_us}) we have $\bm{\nabla} \cdot \bm{\eta}_0
= \varepsilon \partial_U \tilde{\eta}_0^u /\xi + \mathcal{O}(\varepsilon^2)$,
$\nabla^2 = (\partial_U^2 + \varepsilon \tilde{\kappa} \partial_U) / \xi^2
+ \mathcal{O}(\varepsilon^2)$, and $(\nabla^2 + 2i \bm{q}_j^0 \cdot 
\bm{\nabla} )^2 = (\nabla^2 + 2i \bm{q}_j^0 \cdot \bm{\nabla} )_0^2
+ \varepsilon (\nabla^2 + 2i \bm{q}_j^0 \cdot \bm{\nabla} )_1^2
+ \mathcal{O}(\varepsilon^2)$, where for $j=1,3$,
\begin{eqnarray}
\left ( \nabla^2 + 2i \bm{q}_j^0 \cdot \bm{\nabla} \right )_0^2 = &&
\left ( \partial_U^2 - iq_0 \xi \partial_U \right )^2 / \xi^4, \nonumber\\
\left ( \nabla^2 + 2i \bm{q}_j^0 \cdot \bm{\nabla} \right )_1^2 = && \left [
\mp i 2\sqrt{3} q_0 \xi \sin \tilde{\theta} ~ \partial_U 
(\partial_U^2 - iq_0 \xi \partial_U) \right. \nonumber\\
&& + \tilde{\kappa} \partial_U (2\partial_U^2 - 3q_0^2 \xi^2) 
\mp 2\sqrt{3} q_0^2 \xi^2 \partial_U \partial_S  \nonumber\\
&& \left. - 2iq_0 \xi (\tilde{\kappa} \pm \sqrt{3} \partial_S) \partial_U^2 
\right ] / \xi^4, \label{eq:grad13}
\end{eqnarray}
and for $j=2$,
\begin{eqnarray}
\left ( \nabla^2 + 2i \bm{q}_2^0 \cdot \bm{\nabla} \right )_0^2 = &&
\left ( \partial_U^2 + 2iq_0 \xi \partial_U \right )^2 / \xi^4, \nonumber\\
\left ( \nabla^2 + 2i \bm{q}_2^0 \cdot \bm{\nabla} \right )_1^2 = &&
2 \tilde{\kappa} \partial_U^2 (\partial_U + 2iq_0 \xi) / \xi^4.
\label{eq:grad2}
\end{eqnarray}
To address the time relaxation of system in the inner region, as usual
we use a coordinate frame co-moving with the interface at a normal 
velocity $v_n(s) = \varepsilon \xi \tilde{v}(S) + \mathcal{O}(\varepsilon^2)$,
and hence $\partial_t \rightarrow \partial_t - \bm{v} \cdot \bm{\nabla}
= - \varepsilon \tilde{v} \partial_U + \mathcal{O}(\varepsilon^2)$.
The inner expansion of the nonadiabatic amplitude equations 
(\ref{eq:ampl_Aj}) and (\ref{eq:ampl_n0}) can then be given by: 
For $\mathcal{O}(1)$,
\begin{eqnarray}
&\left ( \nabla^2 + 2i \bm{q}_j^0 \cdot \bm{\nabla} \right )_0^2 A_j^0
+ \left. \frac{\partial f}{\partial A_j^*} \right |_0 =0,& \nonumber\\
&\partial_U^2 \left [ \frac{1}{\xi^4} \partial_U^2 \left ( \partial_U^2
+ 2q_0^2 \xi^2 \right ) n_0^0 + \left. \frac{\partial f}{\partial n_0} 
\right |_0 \right ] =0,&
\label{eq:in_0}
\end{eqnarray}
giving the equilibrium chemical potential $\mu_{\text{eq}}
=\mu_0 = \partial_u^2 (\partial_u^2 + 2q_0^2) n_0^0 
+ (\partial f / \partial n_0)|_0$; At $\mathcal{O}(\varepsilon)$,
\begin{eqnarray}
-\tilde{v} \partial_U A_j^0 = && -q_0^2 \left ( \nabla^2 + 2i \bm{q}_j^0 
\cdot \bm{\nabla} \right )_0^2 \tilde{A}_j \nonumber\\
&& -q_0^2 \sum_{k=1}^3 \left ( \left. \frac{\partial^2 f}{\partial A_j^* 
\partial A_k} \right |_0 \tilde{A}_k + \left. \frac{\partial^2 f}
{\partial A_j^* \partial A_k^*} \right |_0 \tilde{A}_k^* \right )
\nonumber\\
&& -q_0^2 \left. \frac{\partial^2 f}{\partial A_j^* \partial n_0} 
\right |_0 \tilde{n}_0 
-q_0^2 \left ( \nabla^2 + 2i \bm{q}_j^0 \cdot \bm{\nabla} \right )_1^2
A_j^0 \nonumber\\
&& - \int_{U}^{U+\bar{\lambda}_y} \frac{dU'}{\bar{\lambda}_y} 
f_{p_j}(A_j^0,n_0^0) ~ e^{\mp i \bar{q}_0 (U'+\tilde{h})} + \tilde{\eta}_j, 
\label{eq:in_1_Aj}
\end{eqnarray}
\begin{eqnarray}
-\tilde{v} \partial_U n_0^0 &=& \frac{1}{\xi^2} \partial_U^2 
\tilde{\mu}_1 + \frac{1}{\xi} \partial_U \tilde{\eta}_0^u, \nonumber\\
&=& \frac{1}{\xi^2} \partial_U^2 \left \{
\frac{2\tilde{\kappa}}{\xi^4} \partial_U \left ( \partial_U^2
+ q_0^2 \xi^2 \right ) n_0^0 \right . \nonumber\\
&& + \left [ \frac{1}{\xi^4} \partial_U^2 \left ( \partial_U^2
+ 2q_0^2 \xi^2 \right ) + \left. \frac{\partial^2 f}{\partial n_0^2} 
\right |_0 \right ] \tilde{n}_0 \nonumber\\
&& \left. + \sum_{j=1}^3 \left ( \left. \frac{\partial^2 f}{\partial n_0 
\partial A_j} \right |_0 \tilde{A}_j
+ \left. \frac{\partial^2 f}{\partial n_0 \partial A_j^*} 
\right |_0 \tilde{A}_j^* \right ) \right \} \nonumber\\
&& + \frac{1}{\xi} \partial_U \tilde{\eta}_0^u, \label{eq:in_1_n0}
\end{eqnarray}
where we have assumed that the nonadiabatic scale coupling effects are
of $\mathcal{O}(\varepsilon)$. For such nonadiabatic term appearing at 
the end of Eq. (\ref{eq:in_1_Aj}), $\bar{\lambda}_y = \lambda_y/\xi$,
$\bar{q}_0=q_0 \xi$, $\tilde{h}=h/\xi$ with $h(x,t)$ the interface height, 
and we have used the transformation
\begin{equation}
u=(y-h) \cos \theta = (y-h)/\sqrt{1+(\partial_x h)^2},
\label{eq:u_h}
\end{equation}
with the lowest-order approximation $u \sim y-h + \mathcal{O}(\varepsilon^2)$
and $dy \sim du + \mathcal{O}(\varepsilon)$.

Multiplying both sides of Eq. (\ref{eq:in_1_Aj}) by $\partial_U {A_j^0}^*$, 
integrating over $\int_{-\bar{\zeta}}^{\bar{\zeta}} dU$ (with 
$\bar{\zeta} = \zeta / \xi \gg 1$), and then adding the results for all 
$j=1,2,3$ and the corresponding complex conjugates, we obtain
\begin{eqnarray}
&& 2\tilde{v} \sum_j \int_{-\bar{\zeta}}^{\bar{\zeta}} dU \left | 
\partial_U A_j^0 \right |^2 =  - \bar{\sigma}_A \tilde{\kappa} 
- p_0 \sin (\bar{q}_0 \tilde{h} + \varphi) \nonumber\\
&& + q_0^2 \sum_j \int_{-\bar{\zeta}}^{\bar{\zeta}} dU \left ( 
\tilde{n}_0 \partial_U {A_j^0}^* - \tilde{A}_j^* \partial_U n_0^0 \right ) 
\left. \frac{\partial^2 f}{\partial A_j^* \partial n_0} \right |_0
+ \text{c.c.} \nonumber\\
&& - \left [ \int_{-\bar{\zeta}}^{\bar{\zeta}} dU \sum_j \tilde{\eta}_j
\partial_U {A_j^0}^* + \text{c.c.} \right ], \label{eq:v_Aj}
\end{eqnarray}
where the boundary conditions $\partial_U^k A_j^0 
(\pm \bar{\zeta}) = 0$ for any order $k$ of derivative and 
$\bar{\zeta} \gg 1$ have been used,
\begin{eqnarray}
\bar{\sigma}_A = && \frac{4q_0^2}{\xi^4} \int_{-\bar{\zeta}}^{\bar{\zeta}} dU
\left \{ \sum_j \left ( \left | \partial_U^2 A_j^0 \right |^2
+ \xi^2 \delta_j \left | \partial_U A_j^0 \right |^2 \right ) \right. 
\nonumber\\
&& +iq_0 \xi \left [ \left ( \partial_U^2 A_1^0 \right ) 
\left ( \partial_U {A_1^0}^* \right ) + \left ( \partial_U^2 A_3^0 \right ) 
\left ( \partial_U {A_3^0}^* \right ) \right. \nonumber\\
&& \left. \left. - 2 \left ( \partial_U^2 A_2^0 \right ) 
\left ( \partial_U {A_2^0}^* \right ) \right ] \right \}
\label{eq:sigma_A}
\end{eqnarray}
with $\delta_1=\delta_3=3q_0^2/2$ and $\delta_2=0$, and
\begin{eqnarray}
p_0 e^{i(\varphi - \frac{\pi}{2})} = && 2 \int du ~ e^{iq_0 u} \left [ 
A_1^0 \partial_u f_{p_1}^*(A_j^0,n_0^0) \right. \nonumber\\
&& \left. + {A_2^0}^* \partial_u f_{p_2}(A_j^0,n_0^0) 
+ A_3^0 \partial_u f_{p_3}^*(A_j^0,n_0^0) \right ]
\label{eq:p0}
\end{eqnarray}
with $p_0>0$. Detailed derivation for this lattice coupling
term can be found in Appendix \ref{append:Aj}.

For Eq. (\ref{eq:in_1_n0}) derived from the conserved dynamics 
of $n_0$, we need to adopt a Green's function method
\cite{re:elder01}. Similarly, two Green's functions are introduced, 
including $G^+(U,S;U',S')$ in the region $0<U<\bar{\zeta}$ with 
surface $S_+$ closed at $S=\pm \infty$, and $G^-(U,S;U',S')$ in the 
region $-\bar{\zeta}<U<0$ with the corresponding surface $S_-$;
they satisfy the equation
\begin{equation}
\partial_U^2 G^{\pm}(U,S;U',S') = \delta(U-U') \delta(S-S'),
\label{eq:G}
\end{equation}
with the boundary conditions $G^{\pm}(U=U'=0)=0$ and 
$\partial_U G^{\pm}(U=\pm \bar{\zeta}) = \partial_{U'} 
G^{\pm}(U'=\pm \bar{\zeta}) = 0$. Multiplying Eq. (\ref{eq:in_1_n0})
by $G^+$ ($G^-$) and integrating over the corresponding region lead to
\begin{eqnarray}
\mp \xi^2 && \int_{0^{\pm}}^{\pm \bar{\zeta}} dU' \oint_{S_{\pm}} dS' 
\tilde{v}(S') ~ G^{\pm}(U,S;U',S') ~ \partial_{U'} n_0^0 \nonumber\\
&& = \tilde{\mu}_1(U,S)
\pm \oint_{S_{\pm}} dS' \left. \left ( G^{\pm} \partial_{U'} 
\tilde{\mu}_1 - \tilde{\mu}_1 \partial_{U'} G^{\pm} \right )
\right |_{U'=0^{\pm}}^{U'=\pm \bar{\zeta}} \nonumber\\
&& \quad \pm \xi \int_{0^{\pm}}^{\pm \bar{\zeta}} dU' \oint_{S_{\pm}} dS'
G^{\pm} \partial_{U'} \tilde{\eta}_0^u.
\label{eq:G_mu}
\end{eqnarray}
Further integrating Eq. (\ref{eq:G_mu}) by $\int dU \partial_U n_0^0$
and using the solutions of $G^{\pm}$ [see Eq. (\ref{eq:G_sol}) 
in Appendix \ref{append:matching}], we find
\begin{eqnarray}
 \tilde{v} \xi^2 && \int_{-\bar{\zeta}}^{\bar{\zeta}} dU \left [ 
n_0^0 - n_0^0(\pm \bar{\zeta}) \right ]^2 = - \Delta n_0^0 \tilde{\mu}_1(0,S)
-\bar{\sigma}_n \tilde{\kappa} \nonumber\\
&& - \sum_j \int_{-\bar{\zeta}}^{\bar{\zeta}} dU \left ( 
\tilde{n}_0 \partial_U {A_j^0}^* - \tilde{A}_j^* \partial_U n_0^0 \right ) 
\left. \frac{\partial^2 f}{\partial A_j^* \partial n_0} \right |_0
+ \text{c.c.} \nonumber\\
&& + (\partial_U \tilde{\mu}_1)_{\pm \bar{\zeta}} \int_{-\bar{\zeta}}^{\bar{\zeta}}
dU \left [ n_0^0 - n_0^0(\pm \bar{\zeta}) \right ] \nonumber\\
&& -\xi \int_{-\bar{\zeta}}^{\bar{\zeta}} dU \left [ n_0^0 - n_0^0(\pm \bar{\zeta})
\right ] \left [ \tilde{\eta}_0^u - \tilde{\eta}_0^u(\pm \bar{\zeta}) \right ],
\label{eq:v_n0}
\end{eqnarray}
where $n_0^0(\pm \bar{\zeta}) = n_0^0(\bar{\zeta})$ for $U>0$ and
$=n_0^0(-\bar{\zeta})$ for $U<0$, $\tilde{\eta}_0^u(\pm \bar{\zeta})
=\tilde{\eta}_0^u(\bar{\zeta})$ for $U>0$ and 
$=\tilde{\eta}_0^u(-\bar{\zeta})$ for $U<0$, the miscibility gap
\begin{equation}
\Delta n_0^0 = n_0^0(\bar{\zeta}) - n_0^0(-\bar{\zeta})
\simeq n_0^0(+\infty) - n_0^0(-\infty)
\label{eq:dn0}
\end{equation}
due to $\bar{\zeta} \gg 1$ in the inner region, and
\begin{equation}
\bar{\sigma}_n = \frac{2}{\xi^4} \int_{-\bar{\zeta}}^{\bar{\zeta}} dU
\left [ \left ( \partial_U^2 n_0^0 \right )^2 - q_0^2 \xi^2
\left ( \partial_U n_0^0 \right )^2 \right ].
\label{eq:sigma_n}
\end{equation}
Also, the integration of Eq. (\ref{eq:in_1_n0}) over $(-\bar{\zeta},
\bar{\zeta})$ yields the conservation condition for the inner solution
\begin{equation}
-\tilde{v} \xi^2 \Delta n_0^0 = \left ( \partial_U \tilde{\mu}_1 
\right )_{\bar{\zeta}} - \left ( \partial_U \tilde{\mu}_1 
\right )_{-\bar{\zeta}} + \xi \left [ \tilde{\eta}_0^u(\bar{\zeta})
- \tilde{\eta}_0^u(-\bar{\zeta}) \right ].
\label{eq:conserv_in}
\end{equation}

Combining Eqs. (\ref{eq:v_Aj}), (\ref{eq:v_n0}), and (\ref{eq:conserv_in}) 
and returning to the original unscaled coordinates $(u,s)$, we obtain the 
following equation governing the normal velocity $v_n$ of the interface
(given $\bar{\zeta} \rightarrow \infty$ for the inner region)
\begin{eqnarray}
v_n && \int_{-\infty}^{+\infty} du \left \{ 2 \sum_j \left | \partial_u A_j^0
\right |^2 + q_0^2 \left [ n_0^0 - n_0^0(\pm \infty) \right ]^2 \right \}
\nonumber\\
&& = - q_0^2 \Delta n_0^0 \mu_1(0,s) - \sigma \kappa - p_0\sin (q_0 h+\varphi)
\nonumber\\
&& - v_n q_0^2 \Delta n_0^0 \int_0^{\infty} du \left [ n_0^0 - n_0^0(+ \infty) 
\right ] + \eta_{\text{in}}, \label{eq:v_in}
\end{eqnarray}
where $\mu_1=\varepsilon \tilde{\mu}_1 = \mu^{\text{in}} - \mu_{\text{eq}}
+ \mathcal{O}(\varepsilon^2)$, the noise $\eta_{\text{in}} = -q_0^2 
\int_{-\infty}^{+\infty} du [n_0^0 - n_0^0(\pm \infty)] \eta_0^u - 
[ \int_{-\infty}^{+\infty} du \sum_j \eta_j \partial_u {A_j^0}^* + 
\text{c.c.}]$, and the surface tension $\sigma$ is determined by 
$\bar{\sigma}_A + q_0^2 \bar{\sigma}_n$, i.e.,
\begin{eqnarray}
\sigma = && 2q_0^2 \int_{-\infty}^{+\infty} du \left \{
2 \sum_j \left ( \left | \partial_u^2 A_j^0 \right |^2
+ \delta_j \left | \partial_u A_j^0 \right |^2 \right ) \right. 
\nonumber\\
&& + 2iq_0 \left [ \left ( \partial_u^2 A_1^0 \right ) 
\left ( \partial_u {A_1^0}^* \right ) + \left ( \partial_u^2 A_3^0 \right ) 
\left ( \partial_u {A_3^0}^* \right ) \right. \nonumber\\
&& \left. \left. - 2 \left ( \partial_u^2 A_2^0 \right ) 
\left ( \partial_u {A_2^0}^* \right ) \right ]
+ \left ( \partial_u^2 n_0^0 \right )^2 - q_0^2
\left ( \partial_u n_0^0 \right )^2 \right \}. \label{eq:sigma}
\end{eqnarray}
Note that to derive Eq. (\ref{eq:v_in}), we have used the condition
\begin{equation}
\int_{-\infty}^{+\infty} du \left [ n_0^0 - n_0^0(\pm \infty)
\right ] = 0 \label{eq:Gibbs_surface}
\end{equation}
for a Gibbs surface to define the interface position $u=0$ 
\cite{re:elder01}. We find that this condition can also be derived at
$\mathcal{O}(\varepsilon)$, as shown in Appendix \ref{append:matching}.

\subsection{Results of interface equations}
\label{sec:result_inteqs}

To \textit{match the inner and outer solutions}, we need to use the
boundary conditions at $u=\pm \zeta$, i.e.,
\begin{eqnarray}
&\mu_1(u=\pm \zeta,s) = \mu_1^{\text{out}}(u=\pm \zeta,s),& \nonumber\\
&\left ( \partial_u \mu_1 \right )_{u=\pm \zeta} = 
\left ( \partial_u \mu_1^{\text{out}} \right )_{u=\pm \zeta},&
\label{eq:matching}
\end{eqnarray}
and carry out the expansion of outer solution $\mu_1^{\text{out}}$ around
the boundary. Based on the derivation given in Appendix 
\ref{append:matching}, from Eqs. (\ref{eq:v_in}) and (\ref{eq:mu_in_out})
we can obtain a modified form of the Gibbs-Thomson relation which 
incorporates the effect of coupling to the underlying lattice
\begin{equation}
\zeta_0 v_n = \lambda - \sigma \kappa - p_0 \sin (q_0 h + \varphi) + \eta_v,
\label{eq:vn}
\end{equation}
where
\begin{equation}
\lambda = -q_0^2 \Delta n_0^0 \delta \mu(0,s)
\label{eq:coeffs}
\end{equation}
with $\delta \mu = \mu^{\text{out}} - \mu_{\text{eq}} (= \varepsilon
\tilde{\mu}_1^{\text{out}})$, and $\Delta n_0^0 = n_0^0(+\infty) - n_0^0(-\infty)$ 
as defined in Eq. (\ref{eq:dn0}), which represents the miscibility gap given by
the difference between bulk equilibrium densities of coexisting liquid and solid 
states. Values of $\Delta n_0^0$ are small but always nonzero below the melting 
point due to the first-order and metastability character of the liquid-solid 
transition. Also, $\zeta_0$ is the kinetic coefficient determined by
\begin{equation}
\zeta_0 = \int_{-\infty}^{+\infty} du \left \{ 2 \sum_j \left |
\partial_u A_j^0 \right |^2 + q_0^2 \left [ {n_0^0}^2 - {n_0^0}^2
(\pm \infty) \right ] \right \}. \label{eq:zeta0}
\end{equation}
The noise term, $\eta_v=\eta_{\text{in}}+\eta_m$ (with $\eta_m$ determined
in Appendix \ref{append:matching}), has zero mean and the correlation
\begin{equation}
\langle \eta_v(s,t) \eta_v(s',t') \rangle = 2D \delta(s-s') \delta(t-t'),
\label{eq:eta_v}
\end{equation}
where $D = \vartheta q_0^2 \Gamma_0 k_BT \zeta_0$, and 
$\vartheta = \vartheta_i = \vartheta_0 =1/7$ as in Eq. (\ref{eq:eta}).

Also the standard continuity condition for interface can be obtained
from Eq. (\ref{eq:conserv_in}) and the matching conditions (see
Appendix \ref{append:matching}), i.e.,
\begin{eqnarray}
v_n \Delta n_0^0 &=& \left. \frac{\partial \delta \mu}{\partial u} 
\right |_{0^-} - \left. \frac{\partial \delta \mu}{\partial u} \right |_{0^+}
\nonumber\\
&=& \left [ \left ( \bm{\nabla} \delta \mu \right )_{\text{solid}}
- \left ( \bm{\nabla} \delta \mu \right )_{\text{liquid}} \right ]
\cdot \hat{n},
\label{eq:conserv_vn}
\end{eqnarray}
where $\bm{\nabla} \delta \mu$ is evaluated at the location of
moving solid-liquid interface. Finally to obtain the chemical potential
deviation $\delta \mu$ at the interface from the outer solution
$\delta A_j = A_j^{\text{out}} - {A_j^0}^{\text{out}}$ and
$\delta n_0 = n_0^{\text{out}} - {n_0^0}^{\text{out}}$, we need the 
1st-order outer equation (\ref{eq:out_1}) which can be rewritten as
\begin{equation}
\left. \frac{\partial f}{\partial A_j^*} \right |_1^{\text{out}}=0, \quad
\partial \delta n_0 / \partial t = \nabla^2 
\left. \frac{\partial f}{\partial n_0} \right |_1^{\text{out}}
= \nabla^2 \delta \mu,
\label{eq:out_Ajn0}
\end{equation}
where ``$|_1^{\text{out}}$'' corresponds to the results of expansion 
up to 1st order of $\delta A_j$ and $\delta n_0$ in the derivatives
$\partial f / \partial A_j^*$ and $\partial f / \partial n_0$.
Note that from the equations $(\partial f / \partial A_j^*)|_1
^{\text{out}}=0$ ($j=1,2,3$), each amplitude $\delta A_j$ can be
expressed as a linear function of $\delta n_0$, and hence 
Eq. (\ref{eq:out_Ajn0}) reduces to a diffusion equation of $\delta n_0$
with the effective diffusion constant depending on ${A_j^0}^{\text{out}}$
and ${n_0^0}^{\text{out}}$.

The combination of Eqs. (\ref{eq:out_Ajn0}), (\ref{eq:vn}), and 
(\ref{eq:conserv_vn}) yields a free-boundary problem, and can be reduced 
to the standard form of sharp-interface equations if we neglect the 
lattice coupling term $p_0 \sin (q_0 h + \varphi)$ in Eq. (\ref{eq:vn}). 
The incorporation of such scale coupling effect is analogous to the case 
of driven sine-Gordon equation describing the roughening properties of 
interface subjected to a periodic pinning potential \cite{re:nozieres87,%
re:hwa91}, or to the case of front locking/pinning in fluid pattern 
formation \cite{re:bensimon88b,re:boyer02,*re:boyer02b}. Given
$v_n=-\partial u / \partial t \simeq \partial h / \partial t
/ [1+(\partial_x h)^2]^{1/2}$ from Eq. (\ref{eq:u_h}) and 
$\kappa = \bm{\nabla} \cdot \hat{n} = - \partial_x^2 h 
/ [1+(\partial_x h)^2]^{3/2}$, for small local surface gradient $\partial_xh$
Eq. (\ref{eq:vn}) can be approximated as
\begin{equation}
\zeta_0 \partial h / \partial t = F_0 + \sigma \partial_x^2 h 
+ \frac{\lambda}{2} \left ( \partial_x h \right )^2 
- p_0 \sin (q_0 h + \varphi) + \eta_v.
\label{eq:h}
\end{equation}
This has the same form as the (1+1)D version of the driven sine-Gordon 
equation introduced by Hwa, Kardar and Paczuski 
\cite{re:hwa91,*re:balibar92,*re:mikheev93,*re:rost94,*re:hwa94}. 
It is a variation of the sine-Gordon equation studied earlier by 
Nozi\`eres and Gallet \cite{re:nozieres87}, with an additional KPZ nonlinear 
term $\lambda (\partial_x h)^2 /2$ \cite{re:kardar86}. Here $F_0 \equiv \lambda$ 
represents a thermodynamic driving force determined by the chemical 
potential difference $\delta \mu$ at the interface $y=h$ (i.e., $u=0$),
and terms $\sigma \partial_x^2 h - p_0 \sin (q_0 h + \varphi)$ can be derived
from the sine-Gordon Hamiltonian. Compared to previous studies, our results
given here in the PFC framework can determine detailed properties of the
important parameters involved (including the kinetic coefficient $\zeta_0$,
surface tension $\sigma$, pinning strength $p_0$, and the driving force $\lambda$),
in particular the explicit dependence on system temperature
and elastic constants. Some example results will be given in the 
next section. However, it is important to note that while the above equation 
(\ref{eq:h}) exhibits as a nonconserved form of interface dynamics, 
it is not complete and should be combined with Eqs. (\ref{eq:out_Ajn0}) 
and (\ref{eq:conserv_vn}) due to the condition of mass conservation required in 
a liquid-solid system.

\section{Applications to crystal layer growth and pinning}
\label{sec:appl}

To illustrate the important effects of nonadiabatic scale coupling on the 
dynamics of interface, we apply the interface equations of motion derived
above to a simplified case of layer-by-layer crystal growth. The results,
in particular the different crystal growth modes of ``continuous'' vs.
``activated'' as well as the temperature and elastic-constant dependence
of lattice pinning effect, can be used for examining the formation and 
evolution of more complicated surface/interface structures or patterns 
in further studies, the details of which will be presented elsewhere.
For simplicity, in the following we consider the long wavelength limit of 
the average density field $n_0$ and hence neglect the gradient terms of $n_0$ 
in Eq. (\ref{eq:F}) for the free energy functional $\mathcal{F}$ (i.e., 
$[(\nabla^2 + q_0^2) n_0]^2 \rightarrow q_0^4 n_0^2$), as such terms usually 
yield higher-order contributions to system properties \cite{re:yeon10}.
The corresponding interface equations of motion given in Sec. \ref{sec:result_inteqs} 
remain unchanged, although in Eq. (\ref{eq:sigma}) for the expression of $\sigma$ 
the gradients of $n_0^0$ can then be neglected.

\subsection{Properties of interface parameters}

One of the most important parameters given in the above derivations is
the strength of interface pinning force $p_0$. As determined by Eq. (\ref{eq:p0}),
it depends on the details of liquid-solid equilibrium profiles $A_j^0$ and
$n_0^0$. These profiles are obtained by numerically solving the 1D 0th-order
amplitude equations given by (\ref{eq:in_0}) in an unscaled form:
$( \nabla^2 + 2i \bm{q}_j^0 \cdot \bm{\nabla} )_0^2 A_j^0
+(\partial f / \partial A_j^*) |_0 =0$, and
$\partial_y^2 (\partial f / \partial n_0) |_0 =0$ for the long wavelength 
limit of $n_0^0$. We use a pseudospectral method in numerical calculations,
and apply the periodic boundary condition by setting the initial configuration as 
2 symmetric liquid-solid interfaces located at $y=L_y/4$ and $3L_y/4$. The 1D
system size $L_y$ perpendicular to the interface is chosen as $L_y=2048 \Delta y$
for all the results shown here, and a numerical grid spacing $\Delta y = 
(2\pi /q_0)/8$ is used.

\begin{figure}
\centerline{\includegraphics[width=0.48\textwidth]{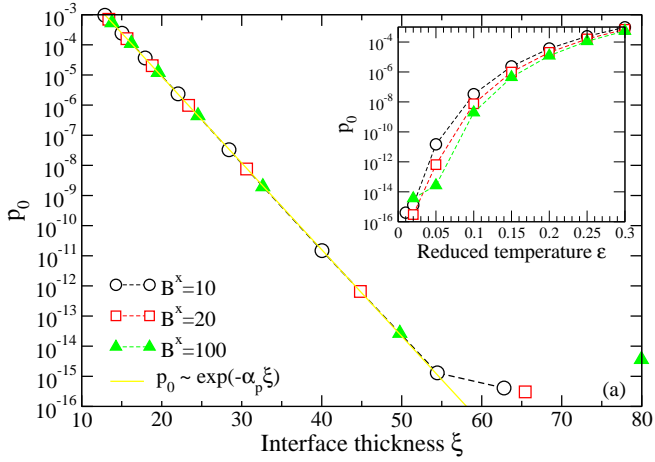}}
\vskip 28pt
\centerline{\includegraphics[width=0.48\textwidth]{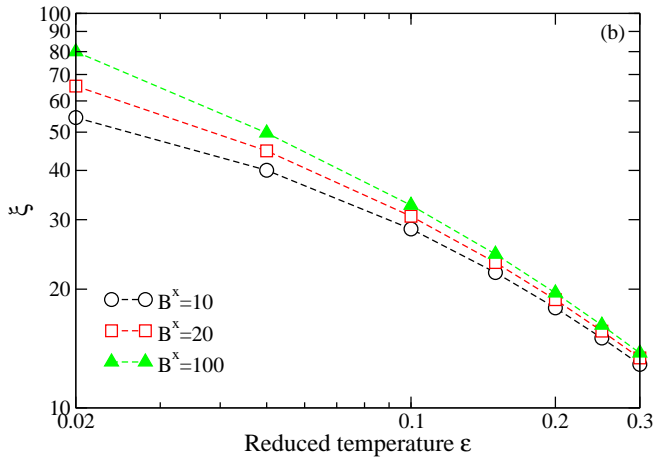}}
\caption{(Color online) (a) Pinning strength $p_0$ as a function of interface 
thickness $\xi$ or reduced temperature $\epsilon$ (inset), for different values 
of elastic modulus $B^x$. The result indicates a universal relation given in 
Eq. (\ref{eq:p0_W}): $p_0 \sim \exp(-\alpha_p \xi)$, with $\alpha_p=0.6620 \pm 
0.0008$. (b) Change of interface thickness $\xi$ with the reduced temperature 
$\epsilon$. Note that larger value of $\epsilon$ corresponds to lower system 
temperature; also for comparison, the numerical value of atomic layer spacing is 
$a_0 = 2\pi /q_0 \sim 6.28$.}
\label{fig:p0}
\end{figure}

As shown in Fig. \ref{fig:p0} (a), the pinning strength $p_0$ increases with the
decrease of system temperature (i.e., with the increasing value of $\epsilon$;
see the inset), and also with the decrease of bulk elastic modulus $B^x$ for
large enough $p_0$ ($>10^{-14}$).
This can be attributed to the phenomenon of sharper liquid-solid interface 
at lower temperature and smaller value of $B^x$ [see Fig. \ref{fig:p0} (b)], 
since sharper interface leads to stronger scale coupling 
between microscopic crystalline structure and mesoscopic amplitudes, and hence
larger pinning force; this is a fundamental mechanism underlying the nonadiabatic 
derivation given in Sec. \ref{sec:ampl}. Thus one would expect that there might 
be a more universal relation between the pinning force and the interface thickness 
$\xi$, as can be derived from Eq. (\ref{eq:p0}) governing $p_0$: Recalling that
both amplitudes $A_j^0$ and $n_0^0$ are functions of scaled variable $U=u/\xi$
in the inner region (see Sec. \ref{sec:inner}), we rewrite Eq. (\ref{eq:p0}) as
\begin{equation}
p_0 = \left | \int_{-\infty}^{+\infty} du e^{iq_0 u} G(u/\xi) \right |,
\label{eq:p0_xi}
\end{equation}
where $G = A_1^0 \partial_u f_{p_1}^* + {A_2^0}^* \partial_u f_{p_2} 
+ A_3^0 \partial_u f_{p_3}^*$. Applying the residue theorem to the integral in
Eq. (\ref{eq:p0_xi}) and assuming that within the poles (singularities) of
$G(U=u/\xi)$ in the upper-half complex plane, the one nearest to the real
axis is given by $U_z = u_z/\xi = \beta_s + i \alpha_s$ (i.e., $\alpha_s$ is
of the smallest value within all poles), we find
\begin{equation}
p_0 \sim e^{-\alpha_p \xi},
\label{eq:p0_W}
\end{equation}
where $\alpha_p = q_0 \alpha_s >0$.
This scaling form is verified in Fig. \ref{fig:p0} (a): All the data 
from different systems characterized by distinct elastic constants 
(i.e., different $B^x$ values) can be scaled onto a single universal 
curve obeying Eq. (\ref{eq:p0_W})
(except for very small values of $p_0 < 10^{-14}$ for which numerical errors 
would be too large), where $\alpha_p=0.6620 \pm 0.0008$ as 
determined from data fitting.

Note that numerical results in Fig. \ref{fig:p0} (a) seem to imply an asymptotic 
behavior of $p_0 \rightarrow 0$ as the system approaches the melting point (i.e., 
$\epsilon \rightarrow 0$). However, since this is a subcritical bifurcation 
system (with hexagonal symmetry), the interface thickness $\xi$ remains 
finite and the pinning strength $p_0$ would never vanish as 
$\epsilon \rightarrow 0$ \cite{re:boyer02b} in both real systems and the full 
PFC model.

\begin{figure}
\centerline{\includegraphics[width=0.48\textwidth]{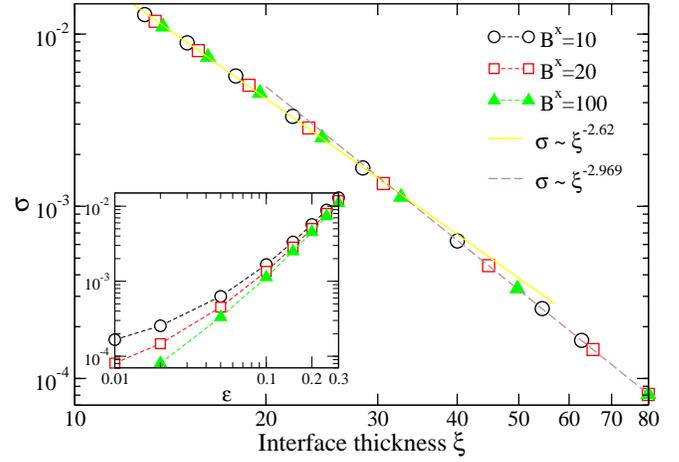}}
\caption{(Color online) Surface tension $\sigma$ as a function of $\xi$ or 
$\epsilon$ (inset). All data for different values of $B^x$ collapses on a single 
universal curve of $\sigma$ vs. $\xi$, which can be fitted into 2 power-law relations:
$\sigma \sim \xi^{-2.62 \pm 0.02}$ for small interface width and $\sigma \sim 
\xi^{-2.969 \pm 0.005}$ for more diffuse interfaces. The crossover occurs at an
intermediate width value $\xi \sim 33$.}
\label{fig:sigma}
\end{figure}

A universal scaling behavior can be also identified for the surface tension
$\sigma$, although the form of scaling is different. As shown in 
Fig. \ref{fig:sigma} where the results are calculated from Eq. (\ref{eq:sigma}),
values of $\sigma$ for systems of different elastic modulus $B^x$ are well 
fitted into a single scaling curve of $\sigma$ vs. $\xi$. This data collapse 
works well for all range of interface thickness $\xi$ in our calculations, 
yielding a power law behavior $\sigma \sim \xi^{-\alpha_{\sigma}}$, although with 
two power-law exponents found in two distinct regimes: For thin enough interface 
$\alpha_{\sigma} = 2.62 \pm 0.02$, while more diffuse interface results in a faster 
decay of $\sigma$ determined by $\alpha_{\sigma} = 2.969 \pm 0.005$. The crossover
between these two scaling regimes is identified in Fig. \ref{fig:sigma}.
Note that generally surface tension $\sigma$ becomes larger for larger value of 
$\epsilon$ (lower temperature), and also for sharper interface with smaller
$B^x$ (see the inset of Fig. \ref{fig:sigma}), as expected.

Another important parameter governing interface motion is the kinetic
coefficient $\zeta_0$, which determines the relation between the interface 
velocity and the thermodynamic driving force and is of large interest 
in solidification studies \cite{re:mikheev91,re:mendelev10,*re:monk10}. 
Note that the expression of $\zeta_0$ given in Eq. (\ref{eq:zeta0}),
if neglecting the last average density term, is similar to that determined by 
Mikheev and Chernov from classical density functional theory \cite{re:mikheev91}
which can well describe recent results of molecular dynamics simulations 
\cite{re:mendelev10}. The anisotropic feature of kinetic coefficient identified
in previous studies has also been incorporated in Eq. (\ref{eq:zeta0}), as the 
amplitude profiles ($A_j^0$ and $n_0^0$) vary with the orientation of liquid-solid 
interface. 

\begin{figure}
\centerline{\includegraphics[width=0.48\textwidth]{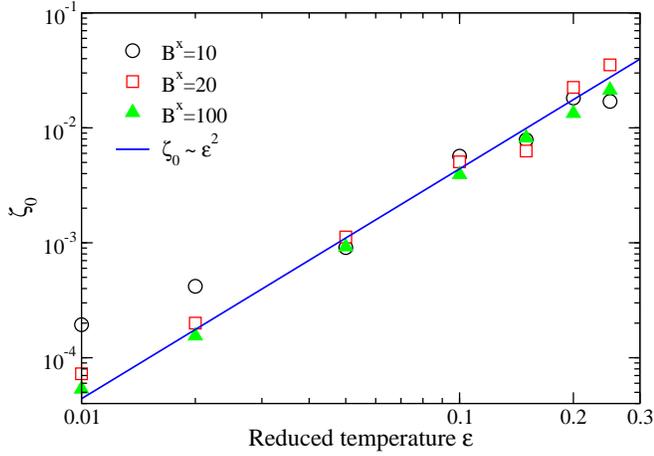}}
\caption{(Color online) Kinetic coefficient $\zeta_0$ as a function of reduced 
temperature $\epsilon$, for different values of $B^x$. An estimate of 
$\zeta_0 \sim \epsilon^2$ is shown for comparison. Note that the thermodynamic 
driving force here is not interface undercooling, but the density supersaturation 
in isothermal systems.}
\label{fig:zeta0}
\end{figure}

On the other hand, here we focus on a system different from these previous 
studies [although the general form of interface equations (\ref{eq:vn}), 
(\ref{eq:out_Ajn0}), and (\ref{eq:conserv_vn}) is applicable to both cases]:
Instead of using interface undercooling as the thermodynamic driving force
\cite{re:mikheev91,re:mendelev10,re:monk10}, here we study the isothermal 
solidification process in pure materials, and the driving force originates
from the supersaturation in atomic density at uniform temperature.
We can then examine the kinetic coefficient for different isothermal
systems, each with a specific $\epsilon$ value and the corresponding 
liquid-solid coexistence conditions. Results evaluated from Eq. (\ref{eq:zeta0})
are given in Fig. \ref{fig:zeta0}, showing the increase of $\zeta_0$ with 
decreasing temperature (i.e., larger $\epsilon$). Data of different elastic
modulus $B^x$ fall around a power-law relation $\zeta_0 \sim \epsilon^2$,
although with large deviation found at small $B^x$ and $\epsilon$ values
(i.e., $B^x=10$ and $\epsilon \leq 0.02$).

\subsection{Planar interface dynamics and pinning}
\label{sec:planar}

In the case of planar growth along the normal $y$ direction, the lateral
variations in the interface equations of motion can be neglected, resulting
in an effective 1D system. In the frame co-moving with the interface at
a velocity $v_0$, the outer equation (\ref{eq:out_Ajn0}) in the liquid
region ($y>0$, and $y \rightarrow y-v_0t$) leads to a steady-state form
\begin{equation}
v_0 \partial_y \delta n_0 + \left ( q_0^4+\beta_0 \right ) 
\partial_y^2 \delta n_0 =0,
\label{eq:dn0_v0}
\end{equation}
where $\beta_0=-\epsilon+3{n_0^0}^2(+\infty)-2gn_0^0(+\infty)$. The
liquid-state chemical potential variation is given by $\delta \mu(y>0) = 
(q_0^4+\beta_0) \delta n_0$, satisfying a far-field boundary condition
$\delta \mu (+\infty) = \Delta$ which represents an external growth 
condition of constant flux coming from the liquid boundary. 
In the solid side ($y<0$), the outer solution yields a 
constant $\delta \mu(y<0) = \delta \mu(0)$. From the interface condition
$v_0 \Delta n_0^0 = - \partial_y \delta \mu |_{0^+}$ as determined 
by Eq. (\ref{eq:conserv_vn}) and the continuity of $\delta \mu$, we obtain
the steady-state solution
\begin{equation}
\delta \mu = \left \{ 
\begin{array}{l} 
\delta \mu_0 \exp \left (-\frac{v_0}{q_0^4+\beta_0} y \right ) + \Delta, 
\quad y \geq 0 \\
\delta \mu_0 + \Delta, \quad y \leq 0,
\end{array} \right.
\end{equation}
where $\delta \mu_0 = (q_0^4+\beta_0) \Delta n_0^0$.
Thus from Eq. (\ref{eq:coeffs}) the effective driving force is given by 
\begin{equation}
F_0 = \lambda = -q_0^2 \Delta n_0^0 (\delta \mu_0 + \Delta).
\label{eq:F0}
\end{equation}

If neglecting the lattice pinning effect, the dynamics of interface profile 
is trivial: $h(t)=h(0)+v_0t$, with a constant interface growth rate 
$v_0 = F_0 / \zeta_0 = -q_0^2 \Delta n_0^0 (\delta \mu_0 + \Delta) / \zeta_0$.
However, as will be shown below the lattice coupling effect plays 
a significant role in the description of interface dynamics, even for 
the simplest case of layer-by-layer growth considered here.

The dynamical equation governing a planar interface profile $h(t)$ is
derived from Eq. (\ref{eq:h}), i.e.,
\begin{equation}
\zeta_0 v_n = \zeta_0 \partial h / \partial t = F_0 
- p_0 \sin (q_0 h + \varphi) + \eta_v, \label{eq:h_1D}
\end{equation}
which can be solved exactly in the absence of the noise term $\eta_v$, 
as given in the following.

\begin{figure}
\centerline{\includegraphics[width=0.48\textwidth]{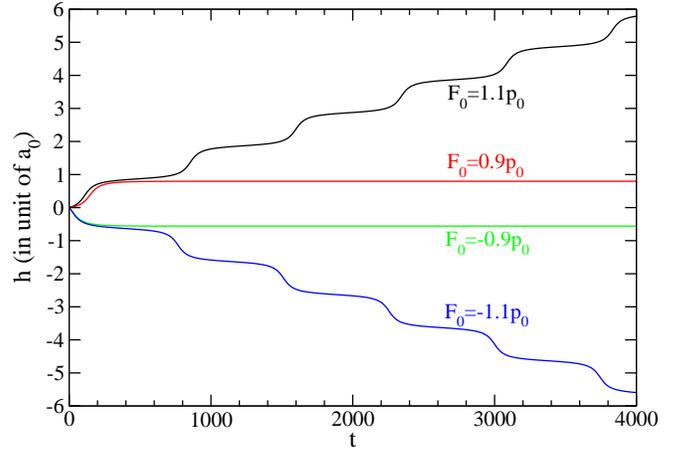}}
\caption{(Color online) Propagation of interface position $h$ with time $t$, 
as evaluated from analytic solutions (\ref{eq:h>}) for $|F_0| = 1.1 p_0$ 
and (\ref{eq:h<}) for $|F_0| = 0.9 p_0$, with $\epsilon =0.3$ 
and $B^x=10$. Positions of $h$ are shown in unit of atomic layer spacing 
$a_0=2\pi /q_0$.}
\label{fig:h_t}
\end{figure}

\subsubsection{$|F_0| > p_0$: continuous growth mode}

When the magnitude of external driving force exceeds the lattice
pinning strength $p_0$, the exact solution of the interface position is
written as
\begin{eqnarray}
h(t) =&& \pm \frac{2}{q_0} \arctan \left \{ \beta \tan \left [
\frac{q_0}{2\zeta_0} \left ( F_0^2-p_0^2 \right )^{1/2} t \pm \tau_0 \right ]
\right \} \nonumber\\
&& - \left ( \varphi - \frac{\pi}{2} \right )/q_0
\label{eq:h>}
\end{eqnarray}
with ``$+$'' for $F_0>0$ and ``$-$'' for $F_0<0$, where 
$\beta=[(F_0-p_0)/(F_0+p_0)]^{1/2}$ and $\tau_0$ is determined by initial 
condition, i.e.,
\begin{equation}
\tau_0 = \arctan \left \{ \frac{1}{\beta} \tan \left [
\frac{q_0}{2} h(0) + \frac{\varphi}{2} - \frac{\pi}{4} \right ] \right \}.
\end{equation}
This situation could occur at small enough $\epsilon$ (i.e., high 
temperature growth) and diffuse enough interface, 
and hence small enough pinning force (see Fig. \ref{fig:p0}), 
given a certain flux condition $F_0$. Despite its form of continuum description, 
the solution (\ref{eq:h>}) yields the jumps of distance $2\pi /q_0$ ($=a_0$)
for interface position, which is exactly one discrete lattice spacing along
the $y$ direction of growth. As shown in Fig. \ref{fig:h_t} which gives the 
numerical evaluation of Eq. (\ref{eq:h>}), the liquid-solid interface propagates 
``continuously'' due to the overcoming of lattice pinning, while the discrete 
lattice effect can still be preserved in this continuous growth
mode as a form of growing steps of $a_0$ spacing. In this case the average 
velocity of interface can be calculated as $\bar{v} = \langle dh/dt \rangle
= \pm (F_0^2 - p_0^2)^{1/2} /\zeta_0$.

\subsubsection{$|F_0| < p_0$: activated/nucleated growth mode}

For the growth condition of lower temperature and sharper interface 
(with larger pinning strength) or weaker driving force such that $|F_0| < p_0$, 
the exact solution of Eq. (\ref{eq:h_1D}) without noise is different:
\begin{equation}
h(t) = \frac{2}{q_0} \arctan \left [ \beta'
\frac{1+\tau_0' f_0(t)}{1-\tau_0' f_0(t)} \right ]
- \left ( \varphi - \frac{\pi}{2} \right )/q_0,
\label{eq:h<}
\end{equation}
where $\beta'=[(p_0-F_0)/(p_0+F_0)]^{1/2}$, 
$f_0(t)=\exp [q_0 (p_0^2-F_0^2)^{1/2} t /\zeta_0]$, and
\begin{equation}
\tau_0' = \frac{\tan \left [ \frac{q_0}{2} h(0) + \frac{\varphi}{2} 
- \frac{\pi}{4} \right ] - \beta'}
{\tan \left [ \frac{q_0}{2} h(0) + \frac{\varphi}{2} 
- \frac{\pi}{4} \right ] + \beta'}.
\end{equation}
The interface growth rate is then given by
\begin{equation}
v_n=dh/dt=\frac{(4/\zeta_0)(p_0-F_0) \tau_0' f_0(t)}
{ \left [ 1-\tau_0' f_0(t) \right ]^2 + \beta'^2 
\left [ 1+\tau_0' f_0(t) \right ]^2 }.
\end{equation}
At large time $t \gg 1$, $v_n=dh/dt \rightarrow 0$; the interface is 
thus locked/pinned by the underlying crystalline potential at a position
$h = -(2\arctan \beta' + \varphi - \pi/2)/q_0$ (satisfying $\sin (q_0h+\varphi)
= F_0/p_0$). This pinning phenomenon is illustrated in Fig. \ref{fig:h_t} 
which shows the numerical evaluation of the analytic solution (\ref{eq:h<}).

Thermal fluctuations should then play an important role on the process
of lattice growth and interface moving, and the full stochastic dynamic 
equation (\ref{eq:h_1D}) with noise term $\eta_v$ governed by 
Eq. (\ref{eq:eta_v}) should be used. This will become a stochastic, 
escape problem in a potential system \cite{re:boyer02}, and the liquid-solid 
front would propagate via an activated process to overcome the pinned lattice 
site, a procedure analogous to thermal nucleation. To illustrate this 
depinning process, we rewrite Eq. (\ref{eq:h_1D}) as 
\begin{equation}
dh / dt = - \frac{\partial U_h}{\partial h} + \eta_h,
\label{eq:h_U}
\end{equation}
where the effective potential $U_h = -[(p_0/q_0) \cos (q_0 h + \varphi) 
+ F_0 h] /\zeta_0$ and the noise $\eta_h$ satisfies 
$\langle \eta_h(t) \eta_h(t') \rangle = 2D_0 \delta(t-t')$, with
$D_0 = D/\zeta_0^2 = \vartheta q_0^2 \Gamma_0 k_BT /\zeta_0$.
From the corresponding Fokker-Planck equation we can determine the
Kramers' escape rate
\begin{equation}
R = \frac{1}{\tau} = \frac{1}{2\pi} \left [ 
\left. \frac{\partial^2 U_h}{\partial h^2} \right |_a
\left | \frac{\partial^2 U_h}{\partial h^2} \right |_b \right ]^2
e^{-\Delta U_h /D_0}, \label{eq:R}
\end{equation}
which represents the rate of an atom hopping/escaping from a metastable
lattice site ``$a$'' (determined as a local minimum of potential $U_h$)
to a nearest lattice site with lower potential, via overcoming
a potential barrier $\Delta U_h = U_h(b) - U_h(a)$ where ``$b$'' indicates 
the top location of the barrier (i.e., a local maximum of $U_h$). In
Eq. (\ref{eq:R}) $\tau$ is the escape time of atoms. It can be shown
that $\partial^2 U_h / \partial h^2 |_a = -\partial^2 U_h / \partial h^2 |_b
= q_0 (p_0^2 - F_0^2)^{1/2} /\zeta_0$, and the potential barrier
\begin{equation}
\Delta U_h = \frac{2}{q_0 \zeta_0} \left [ \left ( p_0^2 - F_0^2 \right )^{1/2}
- |F_0| \arccos \left ( |F_0|/p_0 \right ) \right ] \label{eq:dU}
\end{equation}
(which is always positive for $|F_0| < p_0$).

In this thermally activated process, the lattice nucleation growth rate
is given by $I = a_0 R = a_0 /\tau$, where $a_0=2\pi/q_0$ is the spacing
of atomic layers along the growth direction $\hat{y}$ as shown in
Fig. \ref{fig:hex} and Sec. \ref{sec:ampl}. From Eqs. (\ref{eq:R}),
(\ref{eq:dU}) and the expression of $D_0$, we obtain the standard 
Arrhenius form for thermal nucleation:
\begin{equation}
I = I_0 e^{-E_a/k_BT},
\label{eq:I}
\end{equation}
where
\begin{equation}
I_0 = (p_0^2 - F_0^2)^{1/2} /\zeta_0,
\label{eq:I0}
\end{equation}
and the activation energy $E_a$ is determined by
\begin{equation}
E_a = \frac{2}{\vartheta q_0^3 \Gamma_0}
\left [ \left ( p_0^2 - F_0^2 \right )^{1/2}
- |F_0| \arccos \left ( |F_0|/p_0 \right ) \right ]. \label{eq:Ea}
\end{equation}

\begin{figure}
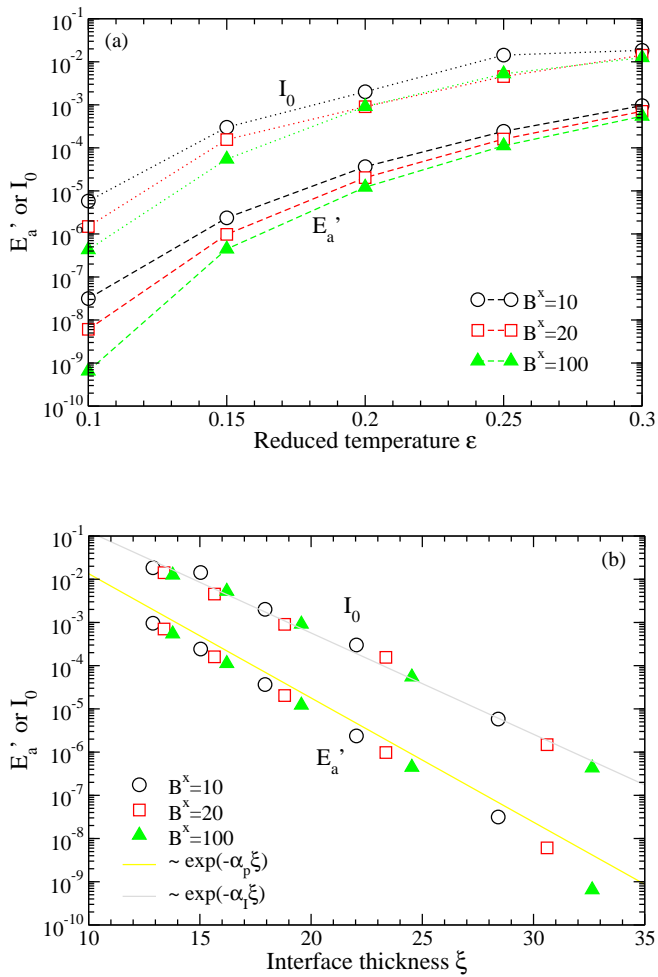

\centerline{\includegraphics[width=0.48\textwidth]{fig6a.eps}}
\vskip 25pt
\centerline{\includegraphics[width=0.48\textwidth]{fig6b.eps}}
\caption{(Color online) The rescaled activation energy $E_a'=E_a \vartheta \Gamma_0$ 
and the prefactor $I_0$ of the nucleation growth rate as a function of (a) reduced
temperature $\epsilon$ and (b) interface width $\xi$, for different values of 
$B^x$ and a constant $F_0=10^{-9}$. In (b), all 
data of $I_0$ are fitted into a scaling form $I_0 \sim \exp (-\alpha_I \xi)$,
where $\alpha_I = 0.54 \pm 0.01$; for the data of $E_a'$, the scaling 
relation of $p_0$ [$\sim \exp (-\alpha_p \xi)$] is shown for comparison.}
\label{fig:Ea_I0}
\end{figure}

It is important to note that in the general form of Eq. (\ref{eq:I}),
both the prefactor $I_0$ and activation energy $E_a$ are actually dependent
on temperature ($\epsilon$) and also elastic constants ($B^x$), as can be 
seen from their expressions in Eqs. (\ref{eq:I0}) and (\ref{eq:Ea}).
For a simple example, if setting a growth condition of $F_0 = \alpha_0 p_0$
($|\alpha_0|<1$) for all temperatures or $\epsilon$ values, we have the
rescaled activation energy $E_a' = E_a \vartheta \Gamma_0 = (2/q_0^3) 
[(1-\alpha_0^2)^{1/2} - |\alpha_0| \arccos |\alpha_0|] p_0$, showing the same
behavior of temperature and interface width dependence as that of $p_0$
(see Fig. \ref{fig:p0}). On the other hand, considering constant driving
force $F_0$ at different temperatures would lead to more complicated 
temperature and width dependence of $E_a$ and $I_0$, as shown in 
Fig. \ref{fig:Ea_I0}. The results there are obtained from 
numerical evaluations of Eqs. (\ref{eq:Ea}) and (\ref{eq:I0}). 
Fig. \ref{fig:Ea_I0} (a) shows that both $E_a'$ and $I_0$ increase with
$\epsilon$ (i.e., the decrease of temperature). At low temperatures 
with large $\epsilon$ and small interface thickness $\xi$,
$p_0 \gg F_0$ and Eq. (\ref{eq:Ea}) yields $E_a' \propto p_0$. We would
then expect the activation energy to follow a universal scaling relation
similar to that of pinning strength $p_0$: $E_a' \sim \exp (-\alpha_p \xi)$.
A deviation would occur for large enough $\xi$ (i.e.,
small enough $\epsilon$ and high enough temperature) due to similar order
of magnitudes between $p_0$ and $F_0$ values, as has been verified in 
Fig. \ref{fig:Ea_I0} (b). Also interestingly, all the numerical data of 
prefactor $I_0$ for different values of elastic modulus ($B^x$) is found
to collapse on a universal curve $I_0 \sim \exp (-\alpha_I \xi)$, where
$\alpha_I = 0.54 \pm 0.01$ as obtained from data fitting.
All these results indicate that in the PFC model the temperature 
dependence of nucleation rate $I$ is not exactly Arrhenius, but 
shows a more complicated nonlinear behavior.

\section{Discussion and Conclusions}

We have constructed a nonadiabatic amplitude representation 
and examined the sharp-interface limit of the 2D single-component PFC model. 
Our main findings include the coupling and interaction between mesoscopic 
description (for ``slow'' variation of structural amplitudes) and microscopic 
scales (for ``fast'' variation of the underlying crystalline structure), and 
also the resulting interface pinning effects which are incorporated in a 
generalized Gibbs-Thomson relation for interface dynamics. The strength of 
the corresponding pinning force, and also the value of solid surface tension,
have been found to obey universal scaling relations with respect to the
liquid-solid interface width. Temperature dependence of the interface
kinetic coefficient has also been examined for isothermal systems driven 
by density supersaturation. The scale coupling effects have 
been illustrated in an example of planar interface growth, which shows
the crossover between two distinct growth modes, a ``continuous'' or
nonactivated mode for high-temperature or strongly-driven growth and 
a ``nucleated'' mode for low/moderate temperature or weakly-driven growth, 
as a result of the competition between the external thermodynamic driving 
force and the lattice pinning/locking effect. Such thermal nucleation process 
in the growth regime of $|F_0|<p_0$ is analogous to a pinning-depinning 
transition in the absence of quenched disorder.

Note that the sample application given in Sec. \ref{sec:planar} can be
viewed as the lowest order approximation of interface growth, and the
scenario of nonactivated/continuous vs. activated growth modes identified is 
qualitatively consistent with the classical crystal growth theory of
Cahn \cite{re:cahn60} based on the concept of critical driving force, 
and also with that found in early work of roughening 
transition by Chui and Weeks \cite{re:chui78} and Nozi\`eres and Gallet 
\cite{re:nozieres87}, although more complicated parameter 
dependence on temperature and material elastic property is determined 
here. More general study should involve nonplanar surface/interface 
evolution and dynamics, so that details of dynamic roughening and faceting 
transition involving this new parameter dependence can be obtained. 
Also, all the above derivations can be readily extended to 
(2+1)D PFC growth systems with 3D bcc or fcc lattice structure. The 
corresponding interface equations are expected to be of the same form as 
Eqs. (\ref{eq:conserv_vn}), (\ref{eq:out_Ajn0}), and the generalized
Gibbs-Thomson relation (\ref{eq:vn}); the latter could be also reduced
to a driven sine-Gordon form (i.e., the form of the (2+1)D Hwa-Kardar-Paczuski 
equation \cite{re:hwa91}) governing the interface profile $h(\bm{r},t)$:
\begin{equation}
\zeta_0 \partial h / \partial t = F_0 + \sigma \nabla^2 h + \frac{\lambda}{2}
\left | \bm{\nabla} h \right |^2 - p_0 \sin (q_0 h + \varphi) + \eta_v,
\label{eq:driven_SG}
\end{equation}
although with more complicated expression of the coefficients. It is important
to note that generally $F_0$ and $\lambda$ are functions of $h$ and lateral
coordinate $\bm{r}$ in both (1+1)D and (2+1)D cases, as can be seen from
$\delta \mu(0,s) = \delta \mu(y=h,x)$ in Eq. (\ref{eq:coeffs}). Thus the 
above driven sine-Gordon form is actually a much more complex
nonlinear equation of $h$, and is closely coupled to the conservation 
condition (\ref{eq:conserv_vn}) and the outer equation (\ref{eq:out_Ajn0})
which reflect the conserved dynamics of atomic number density. On the 
other hand, if neglecting the spatial dependence of $F_0$ and $\lambda$ 
at lowest order approximation [i.e., approximating $F_0$ and $\lambda$ by the 
planar constant result Eq. (\ref{eq:F0}) for the case of weak surface 
fluctuations], the Hwa-Kardar-Paczuski equation with spatially-constant 
coefficients can be recovered. In further studies it would be interesting to 
identify the properties of the corresponding dynamic roughening transition as 
compared to the previous renormalization-group results \cite{re:nozieres87,%
re:hwa91} (noting that from our derivation coefficients in 
Eq. (\ref{eq:driven_SG}) are generally temperature dependent), although one 
would expect such transition to be near the lowest order result of $|F_0|=p_0$.

Another important topic to be addressed, based on the nonadiabatic amplitude 
representation and sharp-interface approach developed here, is the anisotropy 
along different crystal growth directions, particularly for surface tension 
($\sigma$), kinetic coefficient ($\zeta_0$), and pinning strength ($p_0$). 
This would yield detailed properties of facet formation and anisotropic, 
orientation-dependent roughening transition. Note that all the results given 
in Sec. \ref{sec:int_eqs} and \ref{sec:appl} are for interfaces oriented along 
the $y$ direction. For other orientations we expect the same forms of interface 
equations of motion due to similar derivation procedure, but with different
values/expressions of coefficients. For example, as shown explicitly in 
Eqs. (\ref{eq:ampl1})--(\ref{eq:ampl0}) and Fig. \ref{fig:hex}, atomic layers
oriented along $\hat{y}$ or $(\pm \sqrt{3} \hat{x} + \hat{y})/2$ direction
are not equivalent to those along $\hat{x}$ or $(\pm \hat{x} + \sqrt{3} 
\hat{y})/2$ direction, due to different layer spacing ($a_0$ vs. $a_0/\sqrt{3}$)
and hence different degree of scale coupling and pinning effect. Similar
properties are expected for 3D PFC models of various symmetries, although
with more complicated results anticipated.

\begin{acknowledgments}
This work was supported by the National Science Foundation
under Grant No. DMR-0845264.
\end{acknowledgments}

\appendix
\section{Derivation of lattice pinning term in the interface
equation of motion}
\label{append:Aj}

To derive the interface equation from the nonconserved
dynamic equations for $A_j$, we perform
integration of $\int_{-\bar{\zeta}}^{\bar{\zeta}} dU \partial_U {A_j^0}^*$
on Eq. (\ref{eq:in_1_Aj}) and also summation over all the 
resulting equations for $A_j$ and $A_j^*$. Corresponding to the last
scale-coupling term in Eq. (\ref{eq:in_1_Aj}), we get
\begin{widetext}
\begin{eqnarray}
&& - \frac{1}{\bar{\lambda}_y} \sum_j \int_{-\bar{\zeta}}^{\bar{\zeta}} dU
\left [ \left ( \partial_U {A_j^0}^* \right ) \int_{U}^{U+\bar{\lambda}_y} dU' 
f_{p_j}(A_j^0,n_0^0) ~ e^{\mp i \bar{q}_0 (U'+\tilde{h})} + \text{c.c.}
\right ] \nonumber\\
&& = - \frac{e^{i \bar{q}_0 \tilde{h}}}{\bar{\lambda}_y} \left [ 
\int dA_1^0 \int_{U}^{U+\bar{\lambda}_y} dU' f_{p_1}^* ~ e^{i \bar{q}_0 U'}
+ \int d{A_2^0}^* \int_{U}^{U+\bar{\lambda}_y} dU' f_{p_2} ~ e^{i \bar{q}_0 U'}
+ \int dA_3^0 \int_{U}^{U+\bar{\lambda}_y} dU' f_{p_3}^* ~ e^{i \bar{q}_0 U'} 
\right ] + \text{c.c.} \nonumber\\
&& = \frac{e^{i \bar{q}_0 \tilde{h}}}{\bar{\lambda}_y} \left \{
\int dU A_1^0 e^{i \bar{q}_0 U} \left [ f_{p_1}^*(U+\bar{\lambda}_y)
- f_{p_1}^*(U) \right ] + \int dU {A_2^0}^* e^{i \bar{q}_0 U} 
\left [ f_{p_2}(U+\bar{\lambda}_y) - f_{p_2}(U) \right ] \right. \nonumber\\
&& \left. \quad + \int dU A_3^0 e^{i \bar{q}_0 U} \left [ 
f_{p_3}^*(U+\bar{\lambda}_y) - f_{p_3}^*(U) \right ] \right \} 
 + \text{c.c.} \nonumber\\
&& \simeq e^{i \bar{q}_0 \tilde{h}} \int dU ~ e^{i \bar{q}_0 U} \left [ 
A_1^0 \partial_U f_{p_1}^* + {A_2^0}^* \partial_U f_{p_2} 
+ A_3^0 \partial_U f_{p_3}^* \right ] + \text{c.c.},
\label{eq:p0_cos}
\end{eqnarray}
\end{widetext}
where $f_{p_j}(U) \equiv f_{p_j}(A_j^0(U),n_0^0(U))$, and we have used 
$\bar{q}_0 \bar{\lambda}_y = 4\pi$ and $[f_{p_j}(U+\bar{\lambda}_y)  - 
f_{p_j}(U)] / \bar{\lambda}_y \simeq \partial_U f_{p_j}$. It is then
straightforward to show that (\ref{eq:p0_cos}) is equivalent to
the lattice pinning term $p_0 \sin (\bar{q}_0 \tilde{h} + \varphi)$
appearing in Eq. (\ref{eq:v_Aj}), with the pinning strength $p_0$
and phase $\varphi$ determined by Eq. (\ref{eq:p0}).

\section{Matching between inner and outer regions and the Gibbs 
surface condition}
\label{append:matching}

In the inner region, the solution of Eq. (\ref{eq:G}) for Green's 
functions $G^{\pm}$ satisfying the corresponding boundary conditions 
at $U, U'=0, \pm \bar{\zeta}$ has been given in Ref. \cite{re:elder01},
i.e.,
\begin{eqnarray}
&& G^+(U,S; U',S') = \left \{ 
\begin{array}{l} -U' \delta(S-S'), \quad
0 \leq U' < U \leq \bar{\zeta} \\
-U \delta(S-S'), \quad ~0 \leq U < U' \leq \bar{\zeta}
\end{array} \right.
\nonumber\\
&& G^-(U,S; U',S') = \left \{ 
\begin{array}{l} U \delta(S-S'), \quad
-\bar{\zeta} \leq U' < U \leq 0\\
U' \delta(S-S'), \quad -\bar{\zeta} \leq U < U' \leq 0.
\end{array} \right. \nonumber\\
&& \label{eq:G_sol}
\end{eqnarray}
Substituting solution (\ref{eq:G_sol}) into Eq. (\ref{eq:G_mu})
leads to
\begin{eqnarray}
-\tilde{v}(S) \xi^2 && \int_{0^+}^{U} dU' [n_0^0 - n_0^0(+\bar{\zeta})]
= \tilde{\mu}_1(U,S) - \tilde{\mu}_1(0^+,S) \nonumber\\
&& - U \left ( \partial_U \tilde{\mu}_1 \right )_{\bar{\zeta}}
+ \xi \int_{0^+}^{U} dU' [ \tilde{\eta}_0^u - \tilde{\eta}_0^u(+\bar{\zeta})],
\label{eq:mu_G+}\\
-\tilde{v}(S) \xi^2 && \int_{0^-}^{U} dU' [n_0^0 - n_0^0(-\bar{\zeta})]
= \tilde{\mu}_1(U,S) - \tilde{\mu}_1(0^-,S) \nonumber\\
&& - U \left ( \partial_U \tilde{\mu}_1 \right )_{-\bar{\zeta}}
+ \xi \int_{0^-}^{U} dU' [ \tilde{\eta}_0^u - \tilde{\eta}_0^u(-\bar{\zeta})],
\label{eq:mu_G-}
\end{eqnarray}
where Eq. (\ref{eq:mu_G+}) applies to $0^+ \leq U \leq \bar{\zeta}$ and
(\ref{eq:mu_G-}) applies to $-\bar{\zeta} \leq U \leq 0^-$.

From the rescaling $(U,S)=(u/\xi, \varepsilon s/\xi)$ in the inner region
and $(U^o,S^o)=(\varepsilon u/\xi, \varepsilon s/\xi)$ for the outer region,
the matching conditions (\ref{eq:matching}) can be rewritten as
\begin{eqnarray}
&\tilde{\mu}_1(U=\pm \bar{\zeta},S) = \tilde{\mu}_1^{\text{out}}
(U^o=\pm \varepsilon \bar{\zeta},S^o),& \nonumber\\
&\left ( \partial_U \tilde{\mu}_1 \right )_{U=\pm \bar{\zeta}} = \varepsilon
\left ( \partial_{U^o} \tilde{\mu}_1^{\text{out}} \right )_{U^o=\pm \varepsilon 
\bar{\zeta}}.& \label{eq:matching_re}
\end{eqnarray}
Also, due to $\varepsilon \bar{\zeta} = \varepsilon \zeta/\xi \ll 1$
we can carry out the expansion for the outer solution
\begin{equation}
\tilde{\mu}_1^{\text{out}}(0^{\pm},S^o) = \tilde{\mu}_1^{\text{out}}
(\pm \varepsilon \bar{\zeta},S^o) \mp \varepsilon \bar{\zeta}
\left ( \partial_{U^o} \tilde{\mu}_1^{\text{out}} \right )_{\pm \varepsilon 
\bar{\zeta}} + \mathcal{O}(\varepsilon^2). \label{eq:expan_mu}
\end{equation}
Evaluating Eq. (\ref{eq:mu_G+}) with $U=+\bar{\zeta}$ and Eq. (\ref{eq:mu_G-}) 
with $U=-\bar{\zeta}$, and using conditions (\ref{eq:matching_re})
and (\ref{eq:expan_mu}), we find
\begin{eqnarray}
-\tilde{v} \xi^2 \int_{0}^{\bar{\zeta}} dU' && [n_0^0 - n_0^0(+\bar{\zeta})]
= \tilde{\mu}_1^{\text{out}}(0^+,S^o) - \tilde{\mu}_1(0^+,S) \nonumber\\
&& + \xi \int_{0}^{\bar{\zeta}} dU' [ \tilde{\eta}_0^u - 
\tilde{\eta}_0^u(+\bar{\zeta})] + \mathcal{O}(\varepsilon^2),
\label{eq:mu_G+_out}\\
\tilde{v} \xi^2 \int_{-\bar{\zeta}}^0 dU' && [n_0^0 - n_0^0(-\bar{\zeta})]
= \tilde{\mu}_1^{\text{out}}(0^-,S^o) - \tilde{\mu}_1(0^-,S) \nonumber\\
&& - \xi \int_{-\bar{\zeta}}^0 dU' [ \tilde{\eta}_0^u - 
\tilde{\eta}_0^u(-\bar{\zeta})] + \mathcal{O}(\varepsilon^2).
\label{eq:mu_G-_out}
\end{eqnarray}
Adding Eqs. (\ref{eq:mu_G+_out}) and (\ref{eq:mu_G-_out}), using the 
Gibbs surface condition (\ref{eq:Gibbs_surface}), and considering
$\bar{\zeta}=\zeta/\xi \gg 1$, we get
\begin{equation}
\tilde{\mu}_1(0,S) = \tilde{\mu}_1^{\text{out}}(0,S^o)
+ \tilde{v} \xi^2 \int_0^{\infty} dU [n_0^0 - n_0^0(+\infty)]
+ \tilde{\eta}_m,
\label{eq:mu_in_out}
\end{equation}
where $\tilde{\eta}_m = \eta_m/\varepsilon = \{ \int_0^{\infty} du
[\tilde{\eta}_0^u - \tilde{\eta}_0^u(+\infty)] - \int_{-\infty}^0 du
[\tilde{\eta}_0^u - \tilde{\eta}_0^u(-\infty)] \} /2$
(with $\tilde{\eta}_0^u = \eta_0^u/\varepsilon$). Rewriting 
Eq. (\ref{eq:mu_in_out}) in the original scale $(u,s)$ and 
substituting into the interface equation (\ref{eq:v_in}), we can
obtain the generalized Gibbs-Thomson relation given in Eq. (\ref{eq:vn}).

To derive the standard interface continuity condition (\ref{eq:conserv_vn}),
we apply the matching condition (\ref{eq:matching_re}) to Eq. 
(\ref{eq:conserv_in}) and expand the outer result around $u=0$; that is,
$( \partial_U \tilde{\mu}_1 )_{\pm \bar{\zeta}} = \varepsilon
( \partial_{U^o} \tilde{\mu}_1^{\text{out}} )_{\pm \varepsilon \bar{\zeta}}
= \varepsilon ( \partial_{U^o} \tilde{\mu}_1^{\text{out}} )_{0^{\pm}}
\pm \varepsilon^2 \bar{\zeta} ( \partial_{U^o}^2 \tilde{\mu}_1^{\text{out}} 
)_{0^{\pm}} + \mathcal{O}(\varepsilon^3)$. Keeping terms up to 
$\mathcal{O}(\varepsilon)$ and neglecting the noise effect in the outer
solution would then yield Eq. (\ref{eq:conserv_vn}) in the original
scale.

Note that the Gibbs surface condition can actually be determined
from Eqs. (\ref{eq:mu_G+_out}) and (\ref{eq:mu_G-_out}) up to
$\mathcal{O}(\varepsilon)$. Subtracting Eq. (\ref{eq:mu_G+_out})
from (\ref{eq:mu_G-_out}) and neglecting the noise terms, given 
the continuity of $\tilde{\mu}_1$ and $\tilde{\mu}_1^{\text{out}}$ 
at $u=0^{\pm}$ we obtain
\begin{equation}
\int_{-\bar{\zeta}}^{\bar{\zeta}} dU \left [ n_0^0 - n_0^0(\pm \bar{\zeta}) 
\right ] \simeq \mathcal{O}(\varepsilon^2).
\end{equation}
Returning to the original scale and noting $\bar{\zeta} \gg 1$,
at $\mathcal{O}(\varepsilon)$ we can recover Eq. (\ref{eq:Gibbs_surface})
for the Gibbs surface.

\bibliography{../references}

\end{document}